%% file: paper.tex
\documentclass[runningheads]{llncs}
\usepackage{graphicx}
\usepackage{hyperref}

\input{macros}

\lstnewenvironment{boogie}{%
    \lstset{
morekeywords={function,axiom,procedure,int,assert,forall},
keywordstyle=\small\bfseries,
 	    numbersep=5pt,
        basicstyle=\small\ttfamily,
        commentstyle=\color{darkgreen}\small\ttfamily,
        floatplacement={tbp},captionpos=b,
        numbers=none,
        escapechar=~,
        mathescape=true,
        frame=lines,xleftmargin=3pt,xrightmargin=3pt}}{}

\begin{document}
\title{Identifying Overly Restrictive Matching Patterns in SMT-based Program Verifiers
\ifdefined \extendedVersion 
 \\ (extended version)
\else
\fi
}
\titlerunning{Identifying Overly Restrictive Matching Patterns}
%

\author{Alexandra Bugariu \and
Arshavir Ter-Gabrielyan \and
Peter M\"uller}

\authorrunning{A. Bugariu et al.}

%

\institute{Department of Computer Science, ETH Zurich, Switzerland\\
\email{\{alexandra.bugariu,ter-gabrielyan,peter.mueller\}@inf.ethz.ch}}

\maketitle              
\begin{abstract}
\input{abstract}

\end{abstract}

\section{Introduction}\label{sec:introduction}
\input{sections/introduction}

\section{Overview}\label{sec:overview}
\input{sections/overview}

\section{Synthesizing Triggering Terms}  \label{sec:technical}
\input{sections/synthesizing_triggering_terms}

\section{Evaluation}\label{sec:evaluation}
\input{sections/evaluation}

\section{Related Work}\label{sec:related_work}
\input{sections/related_work}

\section{Conclusions}\label{sec:conclusions}
\input{sections/conclusions}

\subsubsection*{Acknowledgements}
We would like to thank the reviewers for their insightful comments. We are also grateful to Felix Wolf for providing us the Gobra benchmarks, and to Evgenii Kotelnikov for his detailed explanations about Vampire.

%
%
%
\newpage
\bibliographystyle{splncs04}
\bibliography{references}

\ifdefined \extendedVersion
\newpage
\begin{subappendices}
\renewcommand{\thesection}{\Alph{section}}
\section{Background: E-matching}\label{sec:background}
\input{sections/background}

\section{Diverse models}\label{sec:diverse_models}
\input{sections/diverse_models}

\section{Extensions}\label{sec:extensions}
\input{sections/extensions}

\section{Additional Examples}\label{sec:more_examples}
\input{sections/additional_examples}

\section{Optimizations}\label{sec:optimizations}
\input{sections/optimizations}

\section{SMT-COMP Benchmarks Selection}\label{sec:benchmarks}
\input{sections/benchmarks_smt_comp}

\end{subappendices}
\else
\fi

\end{document}

%% file: macros.tex
\usepackage{booktabs}   
\usepackage{colortbl}

\usepackage[ruled, noend]{algorithm2e}
\usepackage{etoolbox}

\makeatletter
\patchcmd{\@algocf@start}
  {-1.5em}
  {0.8em}
  {}{}
\makeatother
\makeatletter
\patchcmd\algocf@Vline{\vrule}{\vrule \kern-0.4pt}{}{}
\patchcmd\algocf@Vsline{\vrule}{\vrule \kern-0.4pt}{}{}
\makeatother

\makeatletter
\newcommand{\nosemic}{\renewcommand{\@endalgocfline}{\relax}}
\newcommand{\dosemic}{\renewcommand{\@endalgocfline}{\algocf@endline}}
\let\oldnl\nl
\newcommand{\nonl}{\renewcommand{\nl}{\let\nl\oldnl}}
\makeatother

\SetKwInOut{Arguments}{Arguments}
\SetKwProg{Procedure}{Procedure}{}{}

\SetCommentSty{commentfont}
\SetArgSty{}
\SetFuncArgSty{emph}
\SetProcNameSty{emph}
\SetKw{Break}{break}

\newcommand{\algSynthesizeTriggeringTerm}{\ensuremath{\textsf{synthesizeTriggeringTerm}}}
\newcommand{\algCandidateTerm}{\ensuremath{\textsf{candidateTerm}}}
\newcommand{\algClustersRewritings}{\ensuremath{\textsf{clustersRewritings}}}
\newcommand{\algUnify}{\ensuremath{\textsf{unify}}}
\newcommand{\maxNumModels}{\ensuremath{\mu}}
\newcommand{\maxClusterDepth}{\ensuremath{\delta}}
\newcommand{\inputFormula}{\ensuremath{\texttt{I}}}
\newcommand{\instantiation}[1]{\ensuremath{\textsf{Inst[}#1\textsf{]}}}
\newcommand{\instantiations}[2]{\ensuremath{\textsf{Inst[}#1\textsf{][}#2\textsf{]}}}
\newcommand{\rewritings}[1]{\ensuremath{\textsf{rewritings[}#1\textsf{]}}}

\usepackage[T1]{fontenc}
\usepackage[title]{appendix}
\newcommand{\RN}[1]{%
\scriptscriptstyle{\textup{\expandafter{\romannumeral#1}}}%
}

\usepackage{mathtools}

\usepackage{listings}
\usepackage{xcolor}
\usepackage{bm}
\makeatletter
\newcommand{\shorteq}{%
\settowidth{\@tempdima}{-}
\resizebox{\@tempdima}{\height}{=}%
}
\makeatother

\newcommand*{\SPECSHARP}{Spec\texttt{\#}}
\newcommand*{\SMTCOMP}{SMT\nobreakdash-COMP}

\newcommand{\code}[1]{\text{\lstinline[
  mathescape=true,
	basicstyle=\small\ttfamily,
	morekeywords={var,int}
  ]`#1`}}

\newcommand*{\figref}[1]{Fig.~\ref{#1}}
\newcommand*{\secref}[1]{Sec.~\ref{#1}}
\newcommand*{\tabref}[1]{Tab.~\ref{#1}}
\newcommand*{\algref}[1]{Alg.~\ref{#1}}
\newcommand*{\appendixref}[1]{Appx.~\ref{#1}}

\newcommand{\ie}{{i.e.,\@}}

\newcommand\atgx{\bgroup\markoverwith{\textcolor{violet}{\rule[0.5ex]{2pt}{0.8pt}}}\ULon}

\newcommand*{\thead}[1]{\bfseries #1}

\usepackage{amssymb}

\usepackage{upgreek}

\renewcommand{\sigma}{\upsigma}

\newcommand{\extendedVersion}

%% file: abstract.tex
Universal quantifiers occur frequently in proof obligations produced by program verifiers, for instance, to axiomatize uninterpreted functions and to express properties of arrays. SMT-based verifiers typically reason about them via E-matching, an SMT algorithm that requires syntactic matching patterns to guide the quantifier instantiations. Devising good matching patterns is challenging. In particular, overly restrictive patterns may lead to spurious verification errors if the quantifiers needed for a proof are not instantiated; they may also conceal unsoundness caused by inconsistent axiomatizations. In this paper, we present the first technique that identifies and helps the users remedy the effects of overly restrictive matching patterns. We designed a novel algorithm to synthesize missing triggering terms required to complete a proof. Tool developers can use this information to refine their matching patterns and prevent similar verification errors, or to fix a detected unsoundness.

%% file: sections/introduction.tex
Proof obligations frequently contain universal quantifiers, both in the specification and to encode the semantics of the programming language. Most deductive verifiers \cite{Amighi2016,Astrauskas19,SpecSharp,Shaunak2007,Eilers18,Leino2010,Swamy2013} rely on SMT solvers to discharge the proof obligations via E-matching~\cite{Detlefs2005}. This SMT algorithm requires syntactic matching patterns of ground terms (called \emph{patterns} in the following), to control the instantiations. The pattern $\{\mathtt{f}(x, y)\}$ in the formula 
$\forall x\colon \mathrm{Int}, y\colon \mathrm{Int} :: \{ \mathtt{f}(x, y) \} \;  (x = y) \wedge \neg \mathtt{f}(x, y)$ instructs the solver to instantiate the quantifier \textit{only} when it finds a \emph{triggering term} that matches the pattern, e.g., $\mathtt{f}(7, z)$. The patterns can be written manually or inferred automatically. However, devising them is challenging~\cite{Leino2009,Moskal09}. Too permissive patterns may lead to unnecessary instantiations that slow down verification or even cause non-termination (if each instantiation produces a new triggering term, in a so-called matching loop~\cite{Detlefs2005}). Overly restrictive patterns may prevent the instantiations needed to complete a proof; they cause two major problems in program verification, incompleteness and undetected unsoundness. 

\paragraph{Incompleteness.}

Overly restrictive patterns may cause spurious verification errors when the proof of \emph{valid} proof obligations fails. \figref{fig:motivation-boogie} illustrates this case.
The integer \code{x} represents the address of a node, and the uninterpreted functions \code{len} and \code{nxt} encode operations on linked lists. The axiom defines \code{len}: its result is positive and the last node points to itself. The assertion directly follows from the axiom, but the proof fails because the proof obligation does not contain the triggering term \code{len(nxt(7))}; thus, the axiom does not get instantiated. However, realistic proof obligations often contain  hundreds of quantifiers \cite{SMT-COMP2020}, which makes the manual identification of missing triggering terms extremely difficult.

\begin{figure}[t]
\begin{boogie}
function len(x: int): int;
function nxt(x: int): int;

axiom (forall x: int :: {len(nxt(x))} 
        len(x) > 0 && (nxt(x) == x ==> len(x) == 1) &&
        (nxt(x) != x ==> len(x) == len(nxt(x)) + 1));

procedure trivial() { assert len(7) > 0; }        
\end{boogie}
\vspace{-4mm}
\caption{Example (in Boogie~\cite{boogie}) that leads to a spurious error. The assertion follows from the axiom, but the axiom does not get instantiated without a triggering term.}
\label{fig:motivation-boogie}
\end{figure}

\paragraph{Unsoundness.}

Most of the universal quantifiers in proof obligations appear in axioms over uninterpreted functions (to encode type information, heap models, datatypes, etc.). To obtain sound results, these axioms must be consistent (\ie{}~satisfiable); otherwise all proof obligations hold trivially. Consistency can be proved once and for all by showing the existence of a model, as part of the soundness proof. However, this solution is difficult to apply for those verifiers which generate axioms \emph{dynamically}, depending on the program to be verified. Proving consistency then requires verifying the algorithm that generates the axioms for all possible inputs, and needs to consider many subtle issues~\cite{DarvasLeino07,RudichDarvasMueller08,LeinoMueller08}. 

A more practical approach is to check if the axioms generated for a given program are consistent. However, this check also depends on triggering: an SMT solver may fail to prove unsat if the triggering terms needed to instantiate the contradictory axioms are missing. The unsoundness can thus remain undetected.

\begin{figure}
\begin{small}
\[
\begin{array}{ll}
F_0\colon&  \forall t_0\colon \mathrm{V} :: \{\mathtt{Type}(t_0)\} \; t_0 = \mathtt{ElemType}(\mathtt{Type}(t_0)) \\
F_1\colon & \forall t_1\colon \mathrm{V} :: \{\mathtt{Empty}(t_1)\} \; \mathtt{typ}(\mathtt{Empty}(t_1)) = \mathtt{Type}(t_1) \\
F_2\colon& \forall s_2\colon \mathrm{U}, i_2\colon \mathrm{Int}, v_2\colon \mathrm{U}, l_2\colon \mathrm{Int} :: \{\mathtt{Build}(s_2, i_2, v_2, l_2)\} \\
 &     \hspace{0.7cm} \mathtt{typ}(\mathtt{Build}(s_2, i_2, v_2, l_2)) = \mathtt{Type}(\mathtt{typ}(v_2))\\
F_3\colon &\forall s_3\colon \mathrm{U} :: \{\mathtt{Len}(s_3)\} \; \neg (\mathtt{typ}(s_3) = \mathtt{Type}(\mathtt{ElemType}(\mathtt{typ}(s_3))) \vee  (0 \leq \mathtt{Len}(s_3)) \\
F_4\colon &\forall s_4\colon \mathrm{U}, i_4\colon \mathrm{Int}, v_4\colon \mathrm{U}, l_4\colon \mathrm{Int} :: \{\mathtt{Len}(\mathtt{Build}(s_4, i_4, v_4, l_4))\}  \\
      &\hspace{0.7cm} \neg (\mathtt{typ}(s_4) = \mathtt{Type}(\mathtt{typ}(v_4))) \vee (\mathtt{Len}(\mathtt{Build}(s_4, i_4, v_4, l_4)) = l_4)
\end{array}
\]
\end{small}
\vspace{-4mm}
\caption{Fragment of an old version of Dafny's sequence axiomatization. $\mathrm{U}$ and $\mathrm{V}$ are uninterpreted types. All the named functions are uninterpreted. To improve readability, we use mathematical notation throughout this paper instead of SMT-LIB syntax~\cite{Barrett2017}. 
 \label{fig:Dafny seq}}
\end{figure}

For example, Dafny's~\cite{Leino2010} sequence axiomatization from June~2008 contained an inconsistency found only over a year later. A fragment of this axiomatization is shown in \figref{fig:Dafny seq}. It expresses that empty sequences and sequences obtained through the $\mathtt{Build}$ operation are well-typed ($F_0$--$F_2$), that the length of a type-correct sequence must be non-negative ($F_3$), and that $\mathtt{Build}$ constructs a new sequence of the required length ($F_4$). The intended behavior of $\mathtt{Build}$ is to update the element at index $i_4$ in sequence $s_4$ to $v_4$. However, since there are no constraints on the parameter $l_4$,  $\mathtt{Build}$ can be used with a negative length, leading to a contradiction with $F_3$. This error cannot be detected by checking the satisfiability of the formula $F_0 \wedge \ldots \wedge F_4$, as no axiom gets instantiated.  

\paragraph{This work.} 

For SMT-based deductive verifiers, discharging proof obligations and revealing inconsistencies in axiomatizations require a solver to prove unsat via E-matching. (Verification techniques based on proof assistants are out of scope.) Given an SMT formula for which E-matching yields \emph{unknown} due to insufficient quantifier instantiations, our technique generates suitable triggering terms that allow the solver to complete the proof. These terms enable  users to understand and remedy the revealed completeness or soundness issue. Since the SMT queries for the verification of different input programs are typically very similar, fixing such issues benefits the verification of many or even all future runs of the verifier.

\paragraph{Fixing the incompleteness.} For \figref{fig:motivation-boogie}, our technique finds the triggering term \code{len(nxt(7))}, which allows one to fix the incompleteness. Tool \emph{users} (who cannot change the axioms) can add the term to the program; e.g., adding \code{var t: int;} \code{t := len(nxt(7))} before the assertion has no effect on the execution, but triggers the instantiation of the axiom. Tool \emph{developers} 
can devise less restrictive patterns. For instance, they can move the conjunct \code{len(x) > 0} to a separate axiom with the pattern \code{\{len(x)\}} (simply changing the axiom's pattern to \code{\{len(x)\}}  would cause matching loops). Alternatively, tool developers can adapt the encoding to emit additional triggering terms enforcing certain instantiations~\cite{HeuleKassiosMuellerSummers13,Leino2009}.

\paragraph{Fixing the unsoundness.} In \figref{fig:Dafny seq}, our triggering term $\mathtt{Len}(\mathtt{Build}(\mathtt{Empty}(\mathtt{typ}(v)),$ $0, v, -1))$ (for a fresh value $v$) is sufficient to detect the unsoundness (as shown in 
\ifdefined \extendedVersion 
 \appendixref{sec:background}). 
\else
  Appx. A of \cite{Bugariu21}). 
\fi
Tool developers can use this information to add a precondition to $F_4$, which prevents the construction of sequences with negative lengths. 

\paragraph{Soundness modulo patterns.} \figref{fig:Boogie maps} illustrates another scenario: Boogie's~\cite{boogie} map axiomatization is inconsistent by design at the SMT level \cite{LeinoRummer2010}, but this behavior cannot be exposed from Boogie, as the type system prevents the required instantiations. Thus it does not affect Boogie's soundness. It is nevertheless important to detect it because it could surface if Boogie was extended to support quantifier instantiation algorithms that are not based on E-matching (such as MBQI \cite{MBQI}) or first-order provers. They could \textit{unsoundly} classify an incorrect program that uses this map axiomatization as correct. Since $F_2$ states that storing a key-value pair into a map results in a new map with a potentially \textit{different} type, one can prove that two \textit{different} types (e.g., Boolean and Int) are equal in SMT. This example shows that the problems tackled in this paper cannot be solved by simply switching to other instantiation strategies: these are not the preferred choices of most verifiers \cite{Amighi2016,Astrauskas19,SpecSharp,Shaunak2007,Eilers18,Leino2010,Swamy2013}, and may produce unsound results for verifiers designed for E-matching with axiomatizations sound only modulo patterns.

\begin{figure}[t]
\begin{small}
\[
\begin{array}{ll}
F_0\colon&  \forall kt_0\colon \mathrm{V}, vt_0\colon \mathrm{V}:: \{\mathtt{Type}(kt_0, vt_0)\} \> \mathtt{ValTypeInv}(\mathtt{Type}(kt_0, vt_0)) = vt_0\\
F_1\colon&  \forall m_1\colon \mathrm{U}, k_1\colon \mathrm{U}, v_1\colon \mathrm{U}:: \{\mathtt{Select}(m_1, k_1, v_1)\} \\
& \hspace{0.9cm} \mathtt{typ}(\mathtt{Select}(m_1, k_1, v_1)) = \mathtt{ValTypeInv}(\mathtt{typ}(m_1)) \\
F_2\colon&  \forall m_2\colon \mathrm{U}, k_2\colon \mathrm{U}, x_2\colon \mathrm{U}, v_2\colon \mathrm{U}:: \{\mathtt{Store} (m_2, k_2, x_2, v_2)\} \\
& \hspace{0.9cm} \mathtt{typ}(\mathtt{Store}(m_2, k_2, x_2, v_2)) = \mathtt{Type}(\mathtt{typ}(k_2), \mathtt{typ}(v_2))\\ 
F_3\colon&  \forall m_3\colon \mathrm{U}, k_3\colon \mathrm{U}, x_3\colon \mathrm{U}, v_3\colon \mathrm{U}, k_3'\colon \mathrm{U}, v_3'\colon \mathrm{U}:: \{\mathtt{Select}( \mathtt{Store} (m_3, k_3, x_3, v_3), k_3', v_3')\} \\
& \hspace{0.9cm} (k_3 = k_3') \vee (\mathtt{Select}( \mathtt{Store} (m_3, k_3, x_3, v_3), k_3', v_3') = \mathtt{Select}(m_3, k_3', v_3'))\\ 
\end{array}
\]
\end{small}
\vspace{-4mm}
\caption{Fragment of Boogie's map axiomatization, which is sound only modulo patterns. $\mathrm{U}$ and $\mathrm{V}$ are uninterpreted types. All the named functions are uninterpreted.\label{fig:Boogie maps}}
\end{figure}

\paragraph{Contributions.}

This paper makes the following technical contributions:

\begin{enumerate}
\item We present the first automated technique that allows the developers to detect \emph{completeness} issues in program verifiers and \emph{soundness} problems in their axiomatizations. Moreover, our approach helps them devise better triggering strategies for \textit{all} future runs of their tool with E-matching.

\item We developed a novel algorithm for synthesizing the triggering terms necessary to complete unsatisfiability proofs using E-matching. Since quantifier instantiation is undecidable for first-order formulas over uninterpreted functions, our algorithm might not terminate. However, all identified triggering terms are indeed sufficient to complete the proof; there are no false positives.

\item We evaluated our technique on benchmarks with known triggering problems from four program verifiers. Our experimental results show that it successfully synthesized the missing triggering terms in 65,6\% of the cases, and can significantly reduce the human effort in localizing and fixing the errors.
\end{enumerate}

\paragraph{Outline.}

The rest of the paper is organized as follows: \secref{sec:overview} gives an overview of our technique; the details follow in \secref{sec:technical}.  In \secref{sec:evaluation}, we present our experimental results. We discuss related work in \secref{sec:related_work}, and conclude in \secref{sec:conclusions}. 
\ifdefined \extendedVersion 
\else
  Extensions of our algorithm, optimizations, more details about E-matching and the evaluation, and additional examples can be found in the extended version of our paper \cite{Bugariu21}.
\fi

%% file: sections/overview.tex
\begin{figure}[t]
\includegraphics[scale=0.45,trim=0 220 0 60,clip]{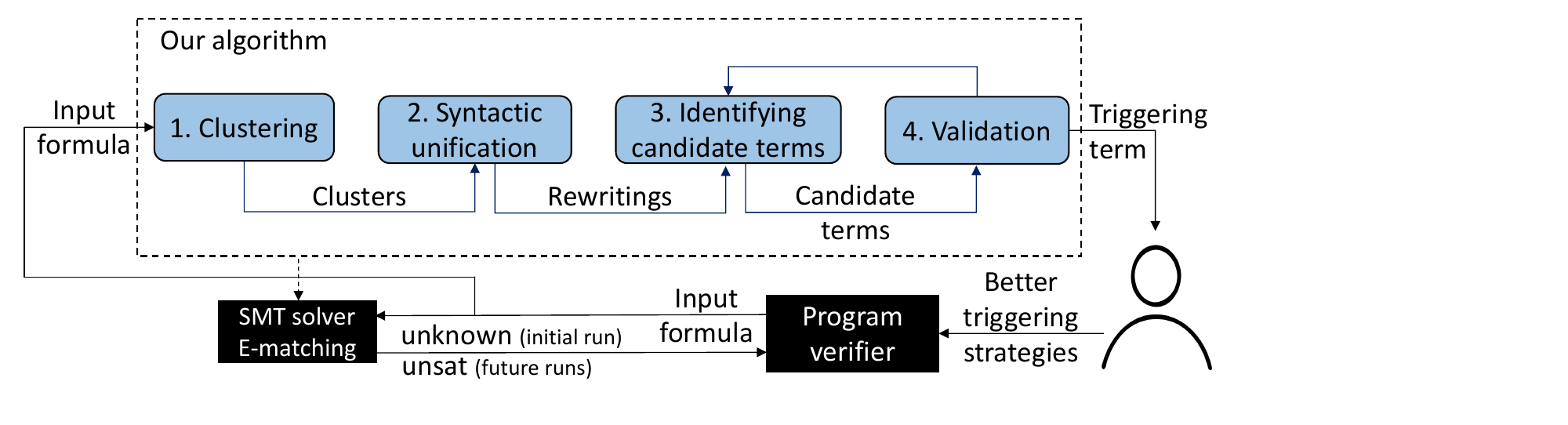}
\vspace{-8mm}
\caption{Main steps of our algorithm that helps the developers of program verifiers devise better triggering strategies. Rounded boxes depict processing steps and arrows data.}
\label{fig:overview}
\end{figure}

Our goal is to synthesize missing triggering terms, i.e., concrete instantiations for (a small subset of) the quantified variables of an input formula \inputFormula, which are necessary for the solver to prove its unsatisfiablity. Intuitively, these triggering terms include \textit{counter-examples} to the satisfiability of \inputFormula $ $ and can be obtained from a model of its negation. For example, $\inputFormula = \forall n\colon \mathrm{Int} :: n > 7$ is unsatisfiable, and a counter-example $n=6$ is a model of its negation $\neg \inputFormula = \exists n\colon \mathrm{Int} :: n \leq 7$.

However, this idea does not apply to formulas over uninterpreted functions, which are common in proof obligations. The negation of $\inputFormula = \exists \mathtt{f}, \forall n\colon \mathrm{Int} :: \mathtt{f}(n, 7)$, where $\mathtt{f}$ is an uninterpreted function, is $\neg \inputFormula =\forall \mathtt{f}, \exists n \colon \mathrm{Int} :: \neg \mathtt{f}(n, 7)$. This is a second-order constraint (it quantifies over functions), and cannot be encoded in SMT, which supports only first-order logic. We thus take a different approach. 

Let $F$ be a second-order formula. We define its \textit{approximation} as:

\begin{equation*}
\label{eqn:approx}
F_{\approx} = F[\exists\overline{\mathtt{f}} \> / \> \forall \overline{\mathtt{f}}]\tag{*}
\end{equation*}

\noindent
where $\overline{\mathtt{f}}$ are uninterpreted functions. The approximation considers only \textit{one} interpretation, not \textit{all} possible interpretations for each uninterpreted function.

We therefore construct a \textit{candidate} triggering term from a model of $\neg \inputFormula_{\approx}$ and check if it is sufficient to prove that $\inputFormula$ is unsatisfiable (due to the approximation, a model is no longer guaranteed to be a counter-example for the original formula).

\medskip
The four main steps of our algorithm are depicted in \figref{fig:overview}. The algorithm is stand-alone, i.e., not integrated into, nor dependent on any specific SMT solver. We illustrate it on the inconsistent axioms from \figref{fig:Inconsistent rewritings} (which we assume are part of a larger axiomatization). To show that $\inputFormula = F_0 \wedge F_1 \wedge \ldots$ is unsatisfiable, the solver requires the triggering term $\mathtt{f}(\mathtt{g}(7))$. The corresponding instantiations of $F_0$ and $F_1$ generate contradictory constraints: $\mathtt{f}(\mathtt{g}(7)) \neq 7$ and $\mathtt{f}(\mathtt{g}(7)) = 7$. In the following, we explain how we obtain this triggering term systematically.

\begin{figure}
\begin{small}
\[
\begin{array}{ll}
F_0\colon\>\> & \forall x_0\colon \mathrm{Int} :: \{\mathtt{f}(x_0)\} \>\>\; \mathtt{f}(x_0) \neq 7 \\
F_1\colon\>\> & \forall x_1\colon \mathrm{Int} :: \{\mathtt{f}(\mathtt{g}(x_1))\} \>\>\; \mathtt{f}(\mathtt{g}(x_1)) = x_1
\end{array}
\]
\end{small}
\vspace{-4mm}
\caption{Formulas that set contradictory constraints on the function $\mathtt{f}$. Synthesizing the triggering term $\mathtt{f}(\mathtt{g}(7))$ requires theory reasoning and syntactic term unification. \label{fig:Inconsistent rewritings}}
\end{figure}

\paragraph{Step 1: Clustering.}

As typical proof obligations or axiomatizations contain hundreds of quantifiers, exploring combinations of triggering terms for all of them does not scale. To prune the search space, we exploit the fact that $\inputFormula$ is unsatisfiable only if there exist instantiations of some (in the worst case all) of its \textit{quantified} conjuncts $F$ such that they produce contradictory constraints on some uninterpreted functions. (If there is a contradiction among the quantifier-free conjuncts, the solver will detect it directly.) We identify \emph{clusters} $C$ of formulas $F$ that share function symbols and then process each cluster separately. In \figref{fig:Inconsistent rewritings}, $F_0$ and $F_1$ share the function symbol $\mathtt{f}$, so we build the cluster $C = F_0 \wedge F_1$.

\paragraph{Step 2: Syntactic unification.}

The formulas within clusters usually contain uninterpreted functions applied to \textit{different} arguments (e.g., $\mathtt{f}$ is applied to $x_0$ in $F_0$ and to $\mathtt{g}(x_1)$ in $F_1$). We thus perform syntactic unification to identify \emph{sharing constraints} on the quantified variables (which we call \textit{rewritings} and denote their set by $R$) such that instantiations that satisfy these rewritings generate formulas with common terms (on which they might set contradictory constraints). $F_0$ and $F_1$ share the term $\mathtt{f}(\mathtt{g}(x_1))$ if we perform the rewritings $R =\{x_0 = \mathtt{g}(x_1)\}$.

\paragraph{Step 3: Identifying candidate triggering terms.}

The cluster $C = F_0 \wedge F_1$ from step~1 contains a contradiction if there exists a formula $F$ in $C$ such that: (1) $F$ is unsatisfiable by itself, or (2) $F$ contradicts at least one other formula from $C$.

To address scenario (1), we ask an SMT solver for a model of the formula $G = \neg C_{\approx}$, where $\neg C_{\approx}$ is defined in \eqref{eqn:approx} above. After Skolemization,  $G$ is quantifier-free, so the solver is generally able to provide a model if one exists. We then obtain a candidate triggering term by substituting the quantified variables from the patterns of the formulas in $C$ with their corresponding values from the model.

However, scenario (1) is not sufficient to expose the contradiction from \figref{fig:Inconsistent rewritings}, since both $F_0$ and $F_1$ are individually satisfiable. Our algorithm thus also derives \emph{stronger} $G$ formulas corresponding to scenario (2). That is, it will next consider the case where $F_0$ contradicts $F_1$, whose encoding into first-order logic is: $\neg {F_0}_{\approx} \wedge F_1 \wedge \bigwedge{R}$, where $R$ is the set of rewritings identified in step~2, used to connect the quantified variables. This formula is  universally-quantified (since $F_1$ is), so the solver cannot prove its satisfiability and generate models. We solve this problem by requiring $F_0$ to contradict the \textit{instantiation} of $F_1$, which is a weaker constraint. Let $F$ be an arbitrary formula. We define its \textit{instantiation} as:

\begin{equation*}
\label{eqn:inst}
F_{Inst} = F[\exists\overline{\mathtt{x}} \> / \> \forall \overline{\mathtt{x}}]\tag{**}
\end{equation*}

\noindent
where $\overline{\mathtt{x}}$ are variables. Then $G=\neg {F_0}_{\approx} \wedge {F_1}_{Inst} \wedge \bigwedge{R}$ is equivalent to $(\mathtt{f}(x_0) = 7) \wedge
(\mathtt{f}(\mathtt{g}(x_1)) = x_1) \wedge
(x_0=\mathtt{g}(x_1))$. (To simplify the notation, here and in the following formulas, we omit existential quantifiers.) All its models set $x_1$ to 7. Substituting $x_0$ by $\mathtt{g}(x_1)$ (according to $R$) and $x_1$ by 7 (its value from the model) in the patterns of $F_0$ and $F_1$ yields the candidate triggering term $\mathtt{f}(\mathtt{g}(7))$.

\paragraph{Step 4: Validation.}

Once we have found a candidate triggering term, we add it to the original formula $\inputFormula$ (wrapped in a fresh uninterpreted function, to make it available to E-matching, but not affect the input's satisfiability) and check if the solver can prove unsat. If so, our algorithm terminates successfully and reports the synthesized triggering term (after a minimization step that removes unnecessary sub-terms); otherwise, we go back to step~3 to obtain another candidate. In our example, the triggering term $\mathtt{f}(\mathtt{g}(7))$ is sufficient to complete the proof.

%% file: sections/synthesizing_triggering_terms.tex
Next, we define the input formulas (\secref{sec:input}), explain the details of our algorithm (\secref{sec:groups}) and discuss its limitations (\secref{sec:limitations}).
\ifdefined \extendedVersion 
  \appendixref{sec:extensions} and \appendixref{sec:optimizations}
\else
  Appx. C and Appx. E of \cite{Bugariu21}
\fi
present extensions that enable complex proofs and optimizations used in \secref{sec:evaluation}.

\subsection{Input formula}\label{sec:input}
\input{sections/input_formula}

\subsection{Algorithm}\label{sec:groups}
\input{sections/groups}

\subsection{Limitations}\label{sec:limitations}
\input{sections/limitations}

%% file: sections/input_formula.tex
To simplify our algorithm, we pre-process the inputs (i.e., the proof obligations or the axioms of a verifier): we Skolemize existential quantifiers and transform all propositional formulas into \emph{negation normal form} (NNF), where negation is applied only to literals and the only logical connectives are conjunction and disjunction; we also apply the distributivity of disjunction over conjunction and split conjunctions into separate formulas. These steps preserve satisfiability and the semantics of patterns
\ifdefined \extendedVersion
  (\appendixref{sec:optimizations} 
\else
  (Appx. E of \cite{Bugariu21} 
\fi
addresses scalability issues). The resulting formulas follow the grammar in \figref{fig:Grammar}.
Literals $L$ may include interpreted and uninterpreted functions, variables and constants. Free variables are nullary functions. Quantified variables can have interpreted or uninterpreted types, and the pre-processing ensures that their names are globally unique. We assume that each quantifier is equipped with a pattern $P$ (if none is provided, we run the solver to infer one). 
Patterns are combinations of \emph{uninterpreted} functions and must mention \emph{all} quantified variables.
Since there are no existential quantifiers after Skolemization, we use the term \emph{quantifier} to denote \emph{universal quantifiers}.

\input{figures/grammar}

%% file: figures/grammar.tex
\begin{figure}[t]
\footnotesize
\[
\begin{array}{@{}llllll@{}}
\inputFormula & ::= & F \; (\wedge \; F)^{*} & \hspace{1cm} B & ::= & D \; (\vee \; D)^{*} \\
F & ::= & B \;|\; \forall \overline{x}\, :: \,\{P(\overline{x})\}\; B & \hspace{1cm}  D & ::= & L \;|\; \neg L \;|\; \forall \overline{x}\, :: \,\{P(\overline{x})\}\; F\\
\end{array}
\]
\vspace{-4mm}
\caption{Grammar of input formulas $\inputFormula$. Inputs are conjunctions of formulas $F$, which are (typically quantified) disjunctions of literals ($L$ or $\neg L$) or nested quantified formulas. Each quantifier is equipped with a pattern $P$. $\overline{x}$ denotes a (non-empty) list of variables.}
\label{fig:Grammar}
\end{figure}


%% file: sections/groups.tex
\begin{algorithm}[t]
\LinesNumbered
\DontPrintSemicolon
\SetFillComment
\SetKwFunction{synthesizeTriggeringTerm}{\algSynthesizeTriggeringTerm}
\SetKwFunction{candidateTerm}{\algCandidateTerm}
\SetKwFunction{clustersRewritings}{\algClustersRewritings}
\SetKwData{None}{None}\SetKwData{Unsat}{UNSAT}\SetKwData{Sat}{SAT}
\SetKwFunction{checkSat}{checkSat}
\SetKwData{InputFormula}{\inputFormula}
\SetKwData{Depth}{depth}
\SetKwData{MaxDepth}{\maxClusterDepth}
\SetKwData{MaxModels}{\maxNumModels}
\BlankLine
\Arguments{\InputFormula~--- input formula, also treated as set of conjuncts\\
           $\sigma$~--- similarity threshold for clustering \\
           \MaxDepth~--- maximum depth for clustering \\
           \MaxModels~--- maximum number of different models}
\KwResult{The synthesized triggering term or \None, if no term was found}
\BlankLine
\Procedure{\synthesizeTriggeringTerm}{
 \BlankLine
  \ForEach{\Depth $\in \{0, \dots, \delta \}$ } {
    \ForEach{$F \in$ \normalfont \InputFormula | $F$ is $ \forall \overline{x} :: F'$} {
      \ForEach(\tcp*[f]{Steps 1,$\:$2}) {$(C, R) \in$ \clustersRewritings{\normalfont\inputFormula, $F,\> \sigma,$ \Depth}} {
        \SetKwArray{Inst}{Inst}
        \BlankLine
        \Inst $\longleftarrow \{\}$ \;
        \ForEach {$f \in C $ | $f$ is $\forall \overline{x} :: D_0 \vee \mathellipsis \vee D_n$ or $D_0 \vee \mathellipsis \vee D_n$} {
          \Inst{$f$} $\longleftarrow\>$ $\{(\bigwedge_{0 \leq j < k} \neg D_j) \wedge D_k$ | $0 \leq k \leq n\}$\;
        }
        \BlankLine
        \Inst{$F$} $\longleftarrow\> \{\neg F'\}$\;
        \BlankLine
        \ForEach(\tcp*[f]{Cartesian product}) {$H \in \bigtimes \{$\Inst{$f$} $| \> f \in \{F\} \cup C\}$} {
          \SetKwData{G}{$G$}
          \G $\longleftarrow\> \bigwedge H \wedge \bigwedge R$\;
          \SetKwData{numModels}{m}
          \SetKwData{candTerm}{$T$}
          \SetKwData{model}{model}
          \SetKwData{resG}{resG}
          \SetKwData{resI}{resI}
          \ForEach{\numModels $\in \{0, \dots, \mu -1\}$} {
            \resG, \model $\longleftarrow$ \checkSat{\G}\;
            \If {\resG $\neq$ \Sat~~} {
            \Break \tcp*{No models if \G is not {\Sat}}} 
              \candTerm $\longleftarrow$ \candidateTerm{$\{F\} \cup C$, $R$, \model} \tcp*{Step 3$\>$}
              \resI, \_ $\longleftarrow$ \checkSat{\InputFormula$\wedge$ \candTerm} \tcp*{Step 4$\>$}
              \If {\resI $=$ \Unsat} {
                \SetKwFunction{minimized}{minimized}
                \Return \minimized{\candTerm} \tcp*{Success$\>$}}
              \G $\longleftarrow$ \G $\wedge \> \neg$\model \tcp*{Avoid this model next iteration$\>$}
         }
       }
      }
    }
  }
  \Return \None
}
\caption{Our algorithm for synthesizing triggering terms that enable unsatisfiability proofs. We assume that all quantified variables are globally unique and \code{I} does not contain nested quantifiers. The auxiliary procedures \code{clustersRewritings} and \code{candidateTerm} are presented in~\algref{lst:Algorithm1} and \algref{lst:Algorithm2}.}
\label{lst:Algorithm}
\end{algorithm}

The pseudo-code of our algorithm is given in \algref{lst:Algorithm}. It takes as input an SMT formula $\inputFormula$ (defined in \figref{fig:Grammar}), which we treat in a slight abuse of notation as both a formula and a set of conjuncts. Three other parameters allow us to customize the search strategy and are discussed later. The algorithm yields a triggering term that enables the unsat proof, or \code{None}, if no term was found. We assume here that $\inputFormula$ contains no nested quantifiers and present those later in this section.
 
The algorithm iterates over each \textit{quantified} conjunct $F$ of $\inputFormula$ (\algref{lst:Algorithm}, line~3) and checks if $F$ is individually unsatisfiable (for \code{depth = 0}). For complex proofs, this is usually not sufficient, as $\inputFormula$ is typically inconsistent due to \emph{a combination} of conjuncts ($F_0 \wedge F_1$ in  \figref{fig:Inconsistent rewritings}). In such cases, the algorithm proceeds as follows:

\input{sections/groups_step1}
\input{sections/groups_step2}

\input{sections/groups_step3}
\input{sections/groups_step4}

\paragraph{Nested quantifiers.} 
Our algorithm also supports nested quantifiers. Nested existential quantifiers in positive positions and nested universal quantifiers in negative positions are replaced in NNF by new, uninterpreted Skolem functions. Step~2 is also applicable to them: Skolem functions with arguments (the quantified variables from the outer scope) are unified as regular uninterpreted functions; they can also appear as $rhs$ in a rewriting, but not as the left-hand side (we do not perform higher-order unification). In such cases, the result is imprecise: the unification of $\mathtt{f}(x_0, \mathtt{skolem}())$ and $\mathtt{f}(x_1, 1)$ produces only the rewriting $x_0 = x_1$.
 
After pre-processing, the conjunct $F$ and the similar formulas may still contain \textit{nested universal quantifiers}. $F$ is always negated in $G$, thus it becomes, after Skolemization, quantifier-free. To ensure that $G$ is also quantifier-free (and the solver can generate a model), we extend the algorithm to \emph{recursively instantiate} similar formulas with nested quantifiers when computing the instantiations.

%% file: sections/groups_step1.tex
\begin{algorithm}[t]
\DontPrintSemicolon
\SetAlgoLined
\LinesNumbered
\SetKwData{InputFormula}{\inputFormula}
\SetKwData{Depth}{depth}
\SetKwData{similarFormulas}{simFormulas}
\SetKwFunction{clustersRewritings}{\algClustersRewritings}
\SetKwFunction{unify}{\algUnify}
\SetKwFunction{qvars}{qvars}
\SetKwFunction{lhs}{lhs}
\SetKwFunction{sim}{sim}
\SetKwData{rws}{rws}
\Arguments{\InputFormula~--- input formula, also treated as set of conjuncts\\
$F$~--- quantified conjunct of \InputFormula, i.e., $F \in $ \InputFormula | $F$ is $\forall \overline{x} :: F'$ \\
$\sigma$~--- similarity threshold for clustering \\
\Depth~--- current depth for clustering  }
\KwResult{A set of pairs, consisting of clusters and their corresponding rewritings}
\BlankLine
\Procedure{\clustersRewritings}{
\If {\Depth = 0} {
\Return $\{(\varnothing, \varnothing)\}$
}
\similarFormulas $\longleftarrow \{f \> | \> f \in \InputFormula  \setminus \{F\} $ and $\sim_{\InputFormula}^{\Depth}$(F, f, $\sigma$)\}
 \tcp*{Step 1}
\SetKwArray{rewritings}{rewritings}
\nl{\rewritings $\longleftarrow \{\}$}\;
\ForEach{$f \in$ \similarFormulas}{
\rws $\longleftarrow$ \unify{$F, f$} \tcp*{Step 2}
\If {\rws $= \varnothing$ and $(f$ is $\forall \overline{x} :: D_0 \vee \mathellipsis \vee D_n)$} {
\similarFormulas $\longleftarrow$ \similarFormulas $\setminus \>\{f\}$
}
\rewritings{$f$} $\longleftarrow$ \rws 
}
\Return \{($C, R$) | $C \subseteq$ \similarFormulas and $(\forall r \in R, \> \exists f \in C: r \in$ \rewritings{$f$}$)$\;
\nonl{\hspace{2.5cm}and $(\forall x \in$ \qvars{C}: $\vert\{r \>| \> r \in R$ and $x =$ \lhs{r}$\}\vert \leq 1) \}$}
}
\caption{Auxiliary procedure for \algref{lst:Algorithm}, which  identifies clusters of formulas similar to $F$ and their rewritings. \code{sim} is defined in text (step~1). \code{unify} is a first-order unification algorithm (not shown); it returns a set of rewritings with restricted shapes, defined in text (step~2).} 
\label{lst:Algorithm1}
\end{algorithm}

\paragraph{Step 1: Clustering.} It constructs clusters of formulas similar to $F$ (\algref{lst:Algorithm1}, line 4), based on their \emph{Jaccard similarity index}. Let $F_i$ and $F_j$ be two arbitrary formulas, and $S_i$ and $S_j$ their respective sets of uninterpreted function symbols (from their bodies and the patterns). The Jaccard similarity index is defined as:

$J(F_i, F_j) =\frac{|S_i \cap S_j|}{|S_i \cup S_j|}$ (the number of common uninterpreted functions divided by the total number). For \figref{fig:Inconsistent rewritings}, $S_0 =\{\mathtt{f}\}$, $S_1 =\{ \mathtt{f, g} \}$, $J(F_0, F_1)  = \frac{|\{\mathtt{f}\}|}{|\{ \mathtt{f, g} \}|} = 0.5$. 

Our algorithm explores the search space by iteratively expanding clusters to include transitively-similar formulas up to a maximum depth (parameter $\maxClusterDepth$ in~\algref{lst:Algorithm}). For two formulas $F_i, F_j \in \inputFormula$, we define the similarity function as:
\[ 
\code{sim}_\inputFormula^{\maxClusterDepth}(F_i, F_j, \sigma)= \left\{
\begin{array}{ll}
      J(F_i, F_j) \geq \sigma, &\> \> \maxClusterDepth = 1 \\
      \exists F_k: \code{sim}_{\inputFormula \setminus \{F_i\}}^{\maxClusterDepth-1}(F_i, F_k, \sigma)$ and $ J(F_k, F_j) \geq \sigma, &\>\>\maxClusterDepth > 1
\end{array} 
\right. 
\]
\noindent
where $\sigma \in [0,1]$ is a similarity threshold used to parameterize our algorithm.

The initial cluster ($\code{depth} = 1$) includes all the conjuncts of $\inputFormula$ that are \textit{directly} similar to $F$. Each subsequent iteration adds the conjuncts that are directly similar to an element of the cluster from the previous iteration, that is, \textit{transitively} similar to $F$. This search strategy allows us to gradually strengthen the formulas $G$ (used to synthesize candidate terms in step 3) without overly constraining them (an over-constrained formula is unsatisfiable, and has no models).

%% file: sections/groups_step2.tex
\paragraph{Step 2: Syntactic unification.} Next (\algref{lst:Algorithm1}, line~8) we identify \emph{rewritings}, \ie{}~constraints under which two similar \textit{quantified} formulas share terms.
\ifdefined \extendedVersion
 (\appendixref{sec:more_examples} 
\else
 (Appx.  D of~\cite{Bugariu21} 
\fi
presents the quantifier-free case.) We obtain the rewritings by performing a \emph{simplified} form of \emph{syntactic term unification}, which reduces their number to a practical size. Our rewritings are \emph{directed equalities}. For two formulas $F_i$ and $F_j$ and an uninterpreted function~$\mathtt{f}$ they have one of the following two shapes:\\
\hspace*{0.5cm}(1) $x_i = \mathit{rhs}_j$, where $x_i$ is a quantified variable of $F_i$, $\mathit{rhs}_j$ are terms from $F_j$ defined below, $F_i$ contains a term $\mathtt{f}(x_i)$ and $F_j$ contains a term $\mathtt{f}(\mathit{rhs}_j)$, \\
\hspace*{0.5cm}(2) $x_j = \mathit{rhs}_i$, where $x_j$ is a quantified variable of $F_j$, $\mathit{rhs}_i$ are terms from $F_i$ defined below, $F_j$ contains a term $\mathtt{f}(x_j)$ and $F_i$ contains a term $\mathtt{f}(\mathit{rhs}_i)$,

\noindent
where $\mathit{rhs}_k$ is a constant $c_k$, a quantified variable $x_k$, or a composite function $(\mathtt{f} \circ \mathtt{g}_0 \circ \ldots \circ \mathtt{g}_n)(\overline{c_k}, \overline{x_k})$ occurring in the formula $F_k$ and $\mathtt{g}_0, \ldots, \mathtt{g}_n$ are arbitrary (interpreted or uninterpreted) functions. That is, we determine the \textit{most general unifier} \cite{Baader2001} only for those terms that have uninterpreted functions as the outer-most functions and quantified variables as arguments. The unification algorithm is standard (except for the restricted shapes), so it is not shown explicitly.

 
Since a term may appear more than once in $F$, or $F$ unifies with multiple similar formulas through the same quantified variable, we can obtain \emph{alternative rewritings} for a quantified variable. In such cases, we either duplicate or split the cluster, such that in each cluster-rewriting pair, each quantified variable is rewritten at most once (see \algref{lst:Algorithm1}, line 12). In \figref{fig:Inconsistent three formulas}, both $F_1$ and $F_2$ are similar to $F_0$ (all three formulas share the uninterpreted symbol $\mathtt{f}$). Since the unification produces alternative rewritings for $x_0$ ($x_0 = x_1$ and $x_0 = x_2$), the procedure \code{clustersRewritings} returns the pairs $\{(\{F_1\}, \{x_0 = x_1\}), (\{F_2\}, \{x_0 = x_2\})\}$. 

\begin{figure}[b]
\begin{small}
\[
\begin{array}{ll}
F_0\colon & \forall x_0\colon \mathrm{Int} :: \{\mathtt{f}(x_0)\} \; \mathtt{f}(x_0) = 6 \\
F_1\colon & \forall x_1\colon \mathrm{Int} :: \{\mathtt{f}(x_1)\} \; \mathtt{f}(x_1) = 7 \\
F_2\colon & \forall x_2\colon \mathrm{Int} :: \{\mathtt{f}(x_2)\} \; \mathtt{f}(x_2) = 8 \\
\end{array}
\]
\end{small}
\vspace{-4mm}
\caption{Formulas that set contradictory constraints on the function $\mathtt{f}$. Synthesizing the triggering term $\mathtt{f}(0)$ requires clusters of similar formulas with alternative rewritings. \label{fig:Inconsistent three formulas}}
\end{figure}

%% file: sections/groups_step3.tex
\paragraph{Step 3: Identifying candidate terms.} From the clusters and the rewritings (identified before), we then derive \emph{quantifier-free} formulas $G$ (\algref{lst:Algorithm}, line 10), and, if they are satisfiable, construct the candidate triggering terms from their models (\algref{lst:Algorithm}, line 15). Each formula $G$ consists of: (1)~$\neg F_{\approx}$ (defined in \eqref{eqn:approx}, which is equivalent to $\neg F'$, since $F$ has the shape $\forall \overline{x} :: F'$ from \algref{lst:Algorithm}, line 3), (2)~the \emph{instantiations} (see \eqref{eqn:inst}) of all the similar formulas from the cluster, and (3)~the corresponding rewritings $R$. (Since we assume that all the quantified variables are globally unique, we do not perform variable renaming for the instantiations).

If a similar formula has multiple disjuncts $D_k$, the solver uses short-circuiting semantics when generating the model for $G$. That is, if it can find a model that satisfies the first disjunct, it does not consider the remaining ones. To obtain more diverse models, we synthesize formulas that \emph{cover} each disjunct, \ie{}~make sure that it evaluates to $\mathtt{true}$ at least once. We thus compute \emph{multiple instantiations} of each similar formula, of the form: $(\bigwedge_{0 \leq j < k} \neg D_j) \wedge D_k, \forall k\colon 0 \leq k \leq n$ (see \algref{lst:Algorithm}, line 7). To consider all the combinations of disjuncts, we derive the formula $G$ from the Cartesian product of the instantiations (\algref{lst:Algorithm}, line 9). (To present the pseudo-code in a concise way, we store $\neg F'$ in the instantiations map as well (\algref{lst:Algorithm}, line 8), even if it does \textit{not} represent the instantiation of $F$.)

In \figref{fig:Inconsistent implications}, $F_1$ is similar to $F_0$ and $R = \{x_0 = x_1\}$. $F_1$ has two disjuncts and thus two possible instantiations: $\instantiation{F_1} = \{x_1 \geq 1, (x_1 < 1) \wedge (\mathtt{f}(x_1) = 6) \}$. The formula $G = (x_0 > -1) \wedge (\mathtt{f}(x_0) \leq 7) \wedge (x_1 \geq 1) \wedge (x_0 = x_1)$ for the first instantiation is satisfiable, but none of the values the solver can assign to $x_0$ (which are all greater or equal to $1$) are sufficient for the unsatisfiability proof to succeed. The second instantiation adds additional constraints: instead of $x_1 \geq 1$, it requires ($x_1 < 1) \wedge (\mathtt{f}(x_1) = 6)$. The resulting $G$ formula has a unique solution for $x_0$, namely 0, and the triggering term $\mathtt{f}(0)$ is sufficient to prove unsat.

\begin{figure}[b]
\begin{small}
\[
\begin{array}{ll}
F_0\colon & \forall x_0\colon \mathrm{Int} :: \{\mathtt{f}(x_0)\} \; \neg (x_0 > -1) \vee (\mathtt{f}(x_0) > 7) \\
F_1\colon & \forall x_1\colon \mathrm{Int} :: \{\mathtt{f}(x_1)\} \; \neg (x_1 <  1) \vee (\mathtt{f}(x_1) = 6) 
\end{array}
\]
\end{small}
\vspace{-4mm}
\caption{Formulas that set contradictory constraints on the function $\mathtt{f}$. Synthesizing the triggering term $\mathtt{f}(0)$ requires instantiations that cover all the disjuncts.}
\label{fig:Inconsistent implications}
\end{figure}

\begin{algorithm}[t]
\DontPrintSemicolon
\SetAlgoLined
\LinesNumbered
\SetKwFunction{candidateTerm}{\algCandidateTerm}
\SetKwData{clusterFormulas}{$C$}
\SetKwData{clusterRewritings}{$R$}
\SetKwFunction{model}{model}
\SetKwFunction{patterns}{patterns}
\SetKwFunction{qvars}{qvars}
\Arguments{\clusterFormulas~--- set of formulas in the cluster \\
           \clusterRewritings~--- set of rewritings for the cluster \\
           \model~--- SMT model, mapping variables to values}
\KwResult{A triggering term with no semantic information}
\BlankLine
\Procedure{\candidateTerm}{
  $P_0 , \mathellipsis , P_k \longleftarrow$ \patterns{\clusterFormulas}\;
  \While {\clusterRewritings$\, \neq \varnothing$} {
    choose and remove $r \longleftarrow (x = rhs)$ from \clusterRewritings\;
    $P_0 , \mathellipsis , P_k \longleftarrow (P_0 , \mathellipsis , P_k)[\>rhs/x\>]$\;
    \clusterRewritings $\longleftarrow$ \clusterRewritings$[\>rhs/x\>]$
  }
   \ForEach{$x \in$ \qvars{C}} {
    $P_0 , \mathellipsis , P_k \longleftarrow (P_0 , \mathellipsis , P_k)[\>$\model{x}$/x\>]$} 
  \Return \texttt{"dummy" + "(" +}  $P_0, \mathellipsis , P_k$  \texttt{+ ")"}
}
\caption{Auxiliary procedure for \algref{lst:Algorithm}, which constructs a triggering term from the given cluster, rewritings, and SMT model. \code{dummy} is a fresh function symbol, which conveys no information about the truth value of the candidate term; thus conjoining it to the input preserves  (un)satisfiability.} 
\label{lst:Algorithm2}
\end{algorithm}

The procedure $\algCandidateTerm$ from \algref{lst:Algorithm2} synthesizes a candidate triggering term $T$ from the model of $G$ and the rewritings $R$. We first collect all the patterns of the formulas from the cluster $C$ (\algref{lst:Algorithm2}, line 2), i.e., of $F$ and of its similar conjuncts (see \algref{lst:Algorithm}, line 15). Then, we \emph{apply} the rewritings, in an arbitrary order (\algref{lst:Algorithm2}, lines 3--6). That is, we substitute the quantified variable $x$ from the left hand side of the rewriting with the right hand side term $rhs$ and propagate this substitution to the remaining rewritings. This step allows us to include in the synthesized triggering terms additional information, which cannot be provided by the solver. Then (\algref{lst:Algorithm2}, lines 7--8) we substitute the remaining variables with their \emph{constant} values from the model (i.e., constants for built-in types, and fresh, unconstrained variables for uninterpreted types).
The resulting triggering term is wrapped in an application to a fresh, uninterpreted function \code{dummy}  to ensure that conjoining it to $\inputFormula$ does not change $\inputFormula$'s satisfiability.

%% file: sections/groups_step4.tex
\paragraph{Step 4: Validation.} We validate the candidate triggering term $T$ by checking if $\inputFormula \wedge T$ is unsatisfiable, i.e., if these particular interpretations for the uninterpreted functions generalize to all interpretations (\algref{lst:Algorithm}, line 16). If this is the case then we return the \textit{minimized} triggering term (\algref{lst:Algorithm}, line 18). The $\mathtt{dummy}$ function has multiple arguments, each of them corresponding to one pattern from the cluster (\algref{lst:Algorithm2}, line 9). This is an over-approximation of the required triggering terms (once instantiated, the formulas may trigger each other), so \code{minimized} removes redundant (sub-)terms. If $T$ does not validate, we re-iterate its construction up to a bound $\maxNumModels$ and strengthen the formula $G$ to obtain a different model (\algref{lst:Algorithm}, lines~19 and 11). 
\ifdefined \extendedVersion
 \appendixref{sec:diverse_models} 
\else
  Appx. B of \cite{Bugariu21}
\fi
discusses heuristics for obtaining \textit{diverse models}.

%% file: sections/limitations.tex
Next, we discuss the limitations of our technique, as well as possible solutions.

\paragraph{Applicability.} Our algorithm effectively addresses a common cause of failed unsatisfiability proofs in program verification, i.e., missing triggering terms. Other causes (e.g., incompleteness in the solver's decision procedures due to undecidable theories) are beyond the scope of our work. Also, our algorithm is tailored to \emph{unsatisfiability} proofs; satisfiability proofs cannot be reduced to unsatisfiability proofs by negating the input, because the negation cannot usually be encoded in SMT (as we have illustrated in \secref{sec:overview}).

\paragraph{SMT solvers.} Our algorithm synthesizes triggering terms as long as the SMT solver can find models for our quantifier-free formulas. However, solvers are incomplete, i.e., they can return \textit{unknown} and generate only \textit{partial models}, which are not guaranteed to be correct. Nonetheless, we also use partial models, as the validation step (step 4 in \figref{fig:overview}) ensures that they do not lead to false positives.

\paragraph{Patterns.}
Since our algorithm is based on patterns (provided or inferred), it will not succeed if they do not permit the necessary instantiations. For example, the formula $\forall x\colon \mathrm{Int}, y\colon \mathrm{Int} :: x = y$ is unsatisfiable. However, the SMT solver cannot automatically infer a pattern from the body of the quantifier, since equality is an interpreted function and must not occur in a pattern. Thus E-matching (and implicitly our algorithm) cannot solve this example, unless the user provides as pattern some uninterpreted function that mentions both $x$ and $y$ (e.g., $\mathtt{f}(x, y)$).

\paragraph{Bounds and rewritings.} Synthesizing triggering terms is generally undecidable. We ensure termination by bounding the search space through various customizable parameters, thus our algorithm misses results not found within these bounds. We also only unify applications of uninterpreted functions, which are common in verification. Efficiently supporting interpreted functions (especially equality) is very challenging for inputs with a small number of types (e.g., from Boogie~\cite{boogie}).

\medskip
Despite these limitations, our algorithm effectively synthesizes the triggering terms required in practical examples, as we experimentally show next.

%% file: sections/evaluation.tex
Evaluating our work requires benchmarks with known triggering issues (i.e., for which E-matching yields \textit{unknown}). Since there is no publicly available suite, in \secref{sec:eval-verifiers} we used manually-collected benchmarks from four verifiers \cite{Leino2010,Swamy2016,Gobra,MuellerSchwerhoffSummers16}. Our algorithm succeeded for 65,6\%. To evaluate its applicability to other verifiers, in \secref{sec:eval-smt-comp} we used \SMTCOMP{}~\cite{SMT-COMP2020} inputs. As they were not designed to expose triggering issues, we developed a filtering step (see 
\ifdefined \extendedVersion
 \appendixref{sec:benchmarks})
\else
  Appx. F of \cite{Bugariu21}) 
\fi
to automatically identify the subset that falls into this category. The results show that our algorithm is suited also for \cite{SpecSharp,VCC,Shaunak2007}. \secref{sec:eval-proofs} illustrates that our triggering terms are simpler than the unsat proofs produced by quantifier instantation and refutation techniques, enabling one to fix the root cause of the revealed issues.

\paragraph{Setup.} We used Z3 (4.8.10)~\cite{Z3} to infer the patterns, generate the models and validate the candidate terms. However, our tool can be used with any solver that supports E-matching and exposes the inferred patterns. We used Z3's \code{NNF} tactic to transform the inputs into NNF and locality-sensitive hashing to compute the clusters. We fixed Z3's random seeds to arbitrary values (\code{sat.random_seed} to 488, \code{smt.random_seed} to 599, and \code{nlsat.seed} to  611). 
We set the (soft) timeout to 600s and the memory limit to 6~GB per run and used a 1s timeout for obtaining a model and for validating a candidate term. The experiments were conducted on a Linux server with 252~GB of~RAM and 32~Intel~Xeon CPUs at 3.3~GHz.

\input{sections/evaluation_verifiers}
\input{sections/evaluation_smt_comp}
\input{sections/evaluation_proofs}

%% file: sections/evaluation_verifiers.tex
\subsection{Effectiveness on verification benchmarks with triggering issues}
\label{sec:eval-verifiers}

\input{tables/verifiers}

First, we used manually-collected benchmarks with known triggering issues from Dafny \cite{Leino2010}, F* \cite{Swamy2016}, Gobra \cite{Gobra}, and Viper \cite{MuellerSchwerhoffSummers16}. 
We reconstructed 4, respectively 2 inconsistent axiomatizations from Dafny and F*, based on the changes from the repositories and the messages from the issue trackers; we obtained 11~inconsistent axiomatizations of arrays and option types from Gobra's developers and collected 15 incompleteness issues from Viper's test suite~\cite{ViperTestSuite}, with at least one assertion needed only for triggering. These contain algorithms for arrays, binomial heaps, binary search trees, and regression tests. The file sizes (minimum-maximum number of formulas or quantifiers) are shown in \tabref{tab:verifiers}, columns 3--4.

\paragraph{Configurations.} We ran our tool with five configurations, to also analyze the impact of its parameters (see \algref{lst:Algorithm} and
\ifdefined \extendedVersion
 \appendixref{sec:extensions}).
\else
 Appx. C of \cite{Bugariu21}). 
\fi
The default configuration C0 has: $\sigma = 0.3$ (similarity threshold), $\beta=64$ (batch size, i.e., the number of candidate terms validated together), $\neg$type (no type-based constraints), $\neg$sub (no unification for sub-terms). The other configurations differ from C0 in the parameters shown in \tabref{tab:verifiers}. All configurations use $\maxClusterDepth = 2$ (maximum transitivity depth), $\maxNumModels = 4$ (maximum number of different models), and 600s timeout per file. 

\paragraph{Results.} Columns 5--9 in \tabref{tab:verifiers} show the number of files solved by each configuration, column 10 summarizes the files solved by at least one. Overall, we found suited triggering terms for 65,6\%, including all F* and Gobra benchmarks. An F* unsoundness exposed by all configurations in $\approx$60s is given in 
\ifdefined \extendedVersion
 \figref{fig:fstar_div}.
\else
 \cite{Bugariu21} (Fig. 9).
\fi
It required two developers to be manually diagnosed based on a bug report~\cite{Fstar1848}. A simplified Gobra axiomatization for option types, solved by C4 in $\approx$13s, is shown in
\ifdefined \extendedVersion
 \figref{fig:Type-based constraints}.
\else
 \cite{Bugariu21} (Fig. 11).
\fi 
Gobra's team spent one week to identify some of the issues. As our triggering terms for F* and Gobra were similar to the manually-written ones, they could have reduced the human effort in localizing and fixing the errors.

Our algorithm synthesized missing triggering terms for 7 Viper files, including the array maximum example~\cite{ArrayMax}, for which E-matching could not prove that the maximal element in a strictly increasing array of size~3 is its last element. Our triggering term \code{loc(a,2)} (\code{loc} maps arrays and integers to heap locations) can be added by a \emph{user} of the verifier to their postcondition. A \emph{developer} can fix the root cause of the incompleteness by including a generalization of the triggering term to arbitrary array sizes: 
\code{len(a)!=0 ==> x==loc(a,len(a)-1).val}. Both result in E-matching refuting the proof obligation in under 0.1s. We also exposed another case where Boogie (used by Viper) is sound only modulo patterns (as in \figref{fig:Boogie maps}).

%% file: tables/verifiers.tex
\begin{table}[t]
\caption{Results on verification  benchmarks with known triggering issues. The columns show: the source of the benchmarks, the number of files (\#), their number of conjuncts ($\#F$) and of quantifiers ($\#\forall$), the number of files for which five configurations (C0--C4) synthesized suited triggering terms, our results across all configurations, the number of unsat proofs generated by Z3 (with MBQI~\cite{MBQI}), CVC4 (with enumerative instantiation~\cite{Reynolds2018}), and Vampire \cite{Kovacs2013} (in CASC mode \cite{Sut16}, using Z3 for ground theory reasoning).}
\label{tab:verifiers}
\setlength{\tabcolsep}{4pt}
\vspace{-4mm}
\begin{center}
\scalebox{0.64}{%
\begin{tabular}{@{}lccccccccc|@{\hskip 0.2cm}c@{\hskip 0cm} c@{\hskip 0cm}c@{}}
\toprule
 & $\bm{\#}$ & $\bm{\#F}$ & $\bm{\#\forall}$ & {\thead C0} & {\thead C1} & {\thead C2} & {\thead C3} & {\thead C4} & {\thead Our} &{\thead Z3} & {\thead CVC4} & {\thead Vampire} \\[-0.3em] 
{\thead Source} & & {\thead min-max} & {\thead min-max} & {\textbf{default}} & {$\bm{\sigma{\shorteq}0.1}$} & {$\bm{\beta{\shorteq}1}$} & {\textbf{type}} & {$\bm{\sigma{\shorteq}0.1\:\wedge\:}$\textbf{sub}} & {\textbf{work}} & {\textbf{MBQI}}& {\textbf{enum inst}} & {\textbf{CASC}$\bm{\:\wedge\:}$\textbf{Z3}}\\
\midrule     

{\thead Dafny} & 4 & 6 - 16 & 5 - 16 & 1 & 1& 1 & 1 & 0 & \textbf{1} & 1 & 0 & 2\\

{\thead F*} & 2 & 18 - 2388 & 15 - 2543 & 1 & 1 & 1 & 1 & 2 & \textbf{2} & 1 & 0 & 2\\     
                                    
{\thead Gobra} & 11 & 64 - 78 & 50 - 63 & 5 & 10 & 1 & 7 & 10 & \textbf{11} & 6 & 0 & 11\\ 

{\thead Viper} & 15 & 84 - 143 & 68 - 203 & 7 & 5 & 3 & 5 & 5 & \textbf{7} & 11 & 0 & 15\\

\\
{\thead Total} & \textbf{32} & & & \multicolumn{5}{c}{\textbf{21 (65,6\%)}} & &  {\textbf{19 (59,3\%)}} & {\hspace{0.4cm}\textbf{0 (0\%)}} & \textbf{30 (93,7\%)} \\
\bottomrule

\multicolumn{13}{l}{$\bm{\sigma}$ = similarity threshold; $\bm{\beta}$ = batch size; \textbf{type} = type-based constraints; \textbf{sub} = sub-terms \hspace{0.2cm}  \textbf{C0}: $\bm{\sigma = 0.3}$; $\bm{\beta = 64}$; $\bm \neg$\textbf{type}; $\neg$\textbf{sub}}
\end{tabular}%
}
\end{center}
\vspace{-5mm}
\end{table}

%% file: sections/evaluation_smt_comp.tex
\subsection{Effectiveness on SMT-COMP benchmarks}
\label{sec:eval-smt-comp}
\input{tables/smt_comp}

Next, we considered 61 SMT-COMP~\cite{SMT-COMP2020} benchmarks from \SPECSHARP~\cite{SpecSharp}, VCC~\cite{VCC}, Havoc~\cite{Shaunak2007}, Simplify~\cite{Detlefs2005}, and the Bit-Width-Independent (BWI) encoding~\cite{Niemetz2019}.

\paragraph{Results.} The results are shown in~\tabref{tab:smt_comp}. Our algorithm enabled E-matching to refute 47.5\% of the files, most of them from Spec\texttt{\#} and VCC/Havoc. We manually inspected some BWI benchmarks (for which the algorithm had worse results) and observed that the validation step times out even with a much higher timeout. This shows that some candidate terms trigger matching loops and explains why C2 (which validates them individually) solved one more file. Extending our algorithm to avoid matching loops, by construction, is left as future work.

%% file: tables/smt_comp.tex
\begin{table}
\caption{Results on SMT-COMP inputs. The columns have the structure from \tabref{tab:verifiers}.
}
\label{tab:smt_comp}
\setlength{\tabcolsep}{4pt}
\vspace{-4mm}
\begin{center}
\scalebox{0.62}{%
\begin{tabular}{@{}lccccccccc|@{\hskip 0.2cm}c@{\hskip 0cm} cc@{}}
\toprule
 & $\bm{\#}$ & $\bm{\#F}$ & $\bm{\#\forall}$ & {\thead C0} & {\thead C1} & {\thead C2} & {\thead C3} & {\thead C4} & {\thead Our} &{\thead Z3} & {\thead CVC4} & {\thead Vampire}\\[-0.3em] 
{\thead Source} & & {\thead min-max} & {\thead min-max} & {\textbf{default}} & {$\bm{\sigma{\shorteq}0.1}$} & {$\bm{\beta{\shorteq}1}$} & {\textbf{type}} & {$\bm{\sigma{\shorteq}0.1\:\wedge\:}$\textbf{sub}} & {\textbf{work}} & {\textbf{MBQI}}& {\textbf{enum inst}} & {\textbf{CASC}$\bm{\:\wedge\:}$\textbf{Z3}}\\
\midrule          
                                    
{\thead \SPECSHARP} & 33 & 28 - 2363 & 25 - 645 & 16 & 16 & 14 & 16 & 15 & \textbf{16} & 16 & 0 & 29\\ 

{\thead VCC/Havoc} & 14 & 129 - 1126 & 100 - 1027 & 11 & 9 & 5 & 11 & 9 & \textbf{11} & 12 & 0 & 14\\

{\thead Simplify} & 1 & 256 & 129 & 0 & 0 & 0 & 0 & 0 & \textbf{0} & 1 & 0 & 0\\ 
    
{\thead BWI} & 13 & 189 - 384 & 198 - 456 & 1 & 1 & 2 & 1 & 1 & \textbf{2} & 12 & 0 & 12\\ 

\\
{\thead Total} & \textbf{61} & & &\multicolumn{5}{c}{\textbf{29 (47,5\%)}}  & & \textbf{41 (67,2\%)} & \textbf{\hspace{0.3cm} 0 (0\%)} & {\textbf{55 (90,1\%)}} \\

\bottomrule

\multicolumn{13}{l}{$\bm{\sigma}$ = similarity threshold; $\bm{\beta}$ = batch size; \textbf{type} = type-based constraints; \textbf{sub} = sub-terms \hspace{0.6cm}  \textbf{C0}: $\bm{\sigma = 0.3}$; $\bm{\beta = 64}$; $\bm \neg$\textbf{type}; $\neg$\textbf{sub}}

\end{tabular}%
}
\end{center}
\vspace{-5mm}
\end{table}

%% file: sections/evaluation_proofs.tex
\subsection{Comparison with unsatisfiability proofs}
\label{sec:eval-proofs}

As an alternative to our work, tool developers could try to \textit{manually} identify triggering issues from refutation proofs, but these do not consider patterns and are harder to understand. Columns 11--13 in \tabref{tab:verifiers} and \tabref{tab:smt_comp} show the number of proofs produced by Z3 with MBQI \cite{MBQI}, CVC4~\cite{Barrett2011} with enumerative instantiation~\cite{Reynolds2018}, and Vampire~\cite{Kovacs2013} using Z3 for ground theory reasoning~\cite{Reger2016} and the CASC \cite{Sut16} portfolio mode with competition presets.
CVC4 failed for all examples (it cannot  
construct proofs for quantified logics), Vampire refuted most of them. Our algorithm outperformed MBQI for F* and Gobra and had similar results for Dafny, Spec\texttt{\#} and VCC/Havoc. All our configurations solved two VCC/Havoc files not solved by MBQI 
\ifdefined \extendedVersion
 (\appendixref{sec:more_examples} 
\else
 (Appx. D of \cite{Bugariu21}
\fi
shows an example). Moreover, our triggering terms are much simpler and directly highlight the root cause of the issues. Compared to our generated term \code{loc(a,2)}, MBQI's proof for Viper's array maximum example has 2135 lines and over 700 reasoning steps, while Vampire's proof has 348 lines and 101 inference steps. Other proofs have similar complexity.

Vampire and MBQI cannot replace our technique: as most deductive verifiers employ E-matching, it is important to help the developers use the algorithm of their choice and return sound results even if they rely on patterns for soundness (as in \figref{fig:Boogie maps}). Our tool can also produce multiple triggering terms (see 
\ifdefined \extendedVersion
  \appendixref{sec:extensions}),
\else
  Appx. C of~\cite{Bugariu21}),
\fi
thus it can reveal \textit{multiple} triggering issues for the same input formula.

%% file: sections/related_work.tex
To our knowledge, no other approach automatically produces the information needed by developers to remedy the effects of overly restrictive patterns. Quantifier instantiation and refutation techniques (discussed next) can produce unsatisfiability proofs, but these are much more complex than our triggering terms.

\paragraph{Quantifier instantiation techniques.} \textit{Model-based quantifier instantiation}~\cite{MBQI} (MBQI) was designed for sat formulas. It checks if the models obtained for the quantifier-free part of the input satisfy the quantifiers, whereas we check if the synthesized triggering terms obtained for some interpretation of the uninterpreted functions generalize to all interpretations. In some cases, MBQI can also generate unsatisfiability proofs, but they require expert knowledge to be understood; our triggering terms are much simpler. \textit{Counterexample-guided quantifier instantiation} \cite{Reynolds2015} is a technique for sat formulas, which synthesizes computable functions from logical specifications. It is applicable to functions whose specifications have explicit syntactic restrictions on the space of possible solutions, which is usually not the case for axiomatizations. Thus the technique cannot directly solve the complementary problem of proving soundness of the axiomatization.

\paragraph{E-matching-based approaches.} R\"ummer~\cite{Rummer2012} proposed a \textit{calculus} for first-order logic modulo linear integer arithmetic that integrates constraint-based free variable reasoning with E-matching. Our algorithm does not require reasoning steps, so it is applicable to formulas from all the logics supported by the SMT solver. \textit{Enumerative instantiation}~\cite{Reynolds2018} is an approach that exhaustively enumerates ground terms from a set of ordered, quantifier-free terms from the input. It can be used to refute formulas with quantifiers, but not to construct proofs (see \secref{sec:eval-proofs}). Our algorithm derives quantifier-free formulas and synthesizes the triggering terms from their models, even if the input does not have a quantifier-free part. It uses also syntactic information to construct complex triggering terms. 

\paragraph{Theorem provers.} First-order theorem provers (e.g., Vampire \cite{Kovacs2013}) also generate refutation proofs. More recent works combine a superposition calculus with theory reasoning \cite{Voronkov2014,Reger2016}, integrating SAT/SMT solvers with theorem provers. We also use unification, but to synthesize triggering terms required by E-matching. However, our triggering terms are much simpler than Vampire's proofs and can be used to improve the triggering strategies for all future runs of the verifier.

%% file: sections/conclusions.tex
We have presented the first automated technique that enables the developers of verifiers remedy the effects of overly restrictive patterns. Since discharging proof obligations and identifying inconsistencies in axiomatizations require an SMT solver to prove the unsatisfiability of a formula via E-matching, we developed a novel algorithm for synthesizing triggering terms that allow the solver to complete the proof. Our approach is effective for a diverse set of verifiers, and can significantly reduce the human effort in localizing and fixing triggering issues.

%% file: sections/background.tex
In this section, we briefly discuss the E-matching-related terminology and explain how this quantifier-instantiation algorithm works on an example.  

\paragraph{Patterns vs triggering terms.} Patterns are syntactic hints attached to quantifiers which instruct the SMT solver when to perform an instantiation. In \figref{fig:Dafny seq}, the quantified formula $F_3$ will be instantiated only when a \textit{triggering term} that matches the \textit{pattern} $\{\mathtt{Len}(s_3)\}$ is encountered during the SMT run (i.e., the triggering term is present in the quantifier-free part of the input formula or is obtained by the solver from the body of a previously-instantiated quantifier).

\paragraph{E-matching.}
We now illustrate how E-matching works on the example from \figref{fig:Dafny seq}; in particular, we show how our synthesized triggering term $\mathtt{Len}(\mathtt{Build}$ $(\mathtt{Empty}(\mathtt{typ}(v)), 0, v, -1)))$ helps the solver to prove unsat when added to the axiomatization ($v$ is a fresh variable of type $\mathrm{U}$). Due to space constraints, we omit unnecessary instantiations. The sub-terms $\mathtt{Empty}(\mathtt{typ}(v))$ and $\mathtt{Len}(\mathtt{Build}$ $(\mathtt{Empty}(\mathtt{typ}(v)), 0, v, -1))$ trigger the instantation of $F_1$ and $F_4$, respectively. The solver obtains the body of the quantifiers for these particular values:
\[
\begin{array}{ll}
B_1\colon & \mathtt{typ}(\mathtt{Empty}(\mathtt{typ}(v))) = \mathtt{Type}(\mathtt{typ}(v))  \\
B_4\colon & \neg(\mathtt{typ}(\mathtt{Empty}(\mathtt{typ}(v))) = \mathtt{Type}(\mathtt{typ}(v))) \: \vee \\
     & (\mathtt{Len}(\mathtt{Build}(\mathtt{Empty}(\mathtt{typ}(v)), 0, v, -1)) = -1) \\
\end{array}
\]

Since the first disjunct of $B_4$ evaluates to $\mathtt{false}$ (from $B_1$), the solver learns that the second disjunct must hold (i.e., the length must be -1); we abbreviate it as $L = -1$. Further, the sub-terms $\mathtt{Build}(\mathtt{Empty}(\mathtt{typ}(v))$ and $\mathtt{Len}(\mathtt{Build}$ $(\mathtt{Empty}(\mathtt{typ}(v)), 0, v, -1))$ of the synthesized triggering term lead to the instantiation of $F_2$ and $F_3$, respectively:
 
\[
\begin{array}{ll}
B_2\colon & \mathtt{type}(\mathtt{Build}(\mathtt{Empty}(\mathtt{typ}(v)), 0, v, -1)) = \mathtt{Type}(\mathtt{typ}(v)) \\
B_3\colon & \neg(\mathtt{typ}(\mathtt{Build}(\mathtt{Empty}(\mathtt{typ}(v)), 0, v, -1)) = \\
      & \mathtt{Type}(\mathtt{ElemType}(\mathtt{typ}(\mathtt{Build}(\mathtt{Empty}(\mathtt{typ}(v)), 0, v, -1))))) \: \vee \\ 
      &  (0 \leq \mathtt{Len}(\mathtt{Build}(\mathtt{Empty}(\mathtt{typ}(v)), 0, v, -1))))
\end{array}
\]

$\mathtt{Type}(\mathtt{ElemType}(\mathtt{typ}(\mathtt{Build}(\mathtt{Empty}(\mathtt{typ}(v)), 0, v, -1))))$ from $B_3$ triggers $F_0$:

\[
\begin{array}{ll}
B_0\colon & \mathtt{ElemType}(\mathtt{typ}(\mathtt{Build}(\mathtt{Empty}(\mathtt{typ}(v)), 0, v, -1))) = \\
    & \mathtt{ElemType}(\mathtt{Type}(\mathtt{ElemType}(\mathtt{typ}(\mathtt{Build}(\mathtt{Empty}(\mathtt{typ}(v)), 0, v, -1)))))
\end{array}
\]
    
By equalizing the arguments of the outer-most $\mathtt{ElemType}$ in $B_0$, the solver learns that the first disjunct of $B_3$ is $\mathtt{false}$. The second disjunct must thus hold (i.e., the length should be positive); we abbreviate it as $0 \leq L$. Since $(L = -1) \wedge (0 \leq L) = \mathtt{false}$, the unsatisfiability proof succeeds.

%% file: sections/diverse_models.tex
In this section, we explain the importance of the parameter $\maxNumModels$ from \algref{lst:Algorithm} (the maximum number of models) and discuss heuristics for obtaining diverse models.

\begin{figure}[t]
\begin{small}
\[
\begin{array}{ll}
F\colon & \forall x\colon \mathrm{Int}, y\colon \mathrm{Int}:: \{\mathtt{\_div}(x, y)\} \; \mathtt{\_div}(x, y) = x / y  \\
\end{array}
\]
\end{small}
\vspace{-4mm}
\caption{Inconsistent axiom from F*~\cite{Swamy2016}. $\mathtt{\_div}\colon \mathrm{Int}  \times \mathrm{Int} \rightarrow \mathrm{Int}$ is an uninterpreted function. Synthesizing the triggering term $\mathtt{dummy}(\mathtt{\_div}(1, 2))$ requires diverse models. \label{fig:fstar_div}}
\end{figure}

\medskip
Let us consider the formula from  \figref{fig:fstar_div}, which was part of an  axiomatization with 2,495 axioms. $F$ axiomatizes the uninterpreted function $\mathtt{\_div}\colon \mathrm{Int}  \times \mathrm{Int} \rightarrow \mathrm{Int}$ and is inconsistent, because there exist two integers whose real division ("/") is not an integer. The model produced by the solver for the formula $G = \neg F'$ is $x = -1, y = 0$. $-1/0$ is defined ("/" is a total function \cite{Barrett2017}), but its result is not specified. Thus the solver cannot validate this model (i.e., it returns \textit{unknown}).

In such cases or when the candidate term does not generalize to all interpretations of the uninterpreted functions, we re-iterate its construction, up to the bound $\maxNumModels$ (\algref{lst:Algorithm}, line~11). For this, we strengthen the previously-derived formula $G$ to force the solver find a different model. In \figref{fig:fstar_div}, if we simply exclude previous models, we can obtain a sequence of models with different values for the numerator, but with the same value (0) for the denominator. There are infinitely many such models, and all of them fail to validate for the same reason. 

There are various heuristics one can employ to guide the solver's search for a new model and our algorithm can be parameterized with different ones. In our experiments, we interpret the conjunct $\neg\mathtt{model}$ from \algref{lst:Algorithm}, line 19 as 
$\bigwedge_{x\in\overline{x}}  x \neq\mathtt{model}(x) \wedge \bigwedge_{x_i, x_j\in\overline{x}, \; i \neq j, \; \mathtt{model}(x_i) = \mathtt{model}(x_j)} x_i \neq x_j)$. The first component requires all the variables to have different values than before. This requirement may be too strong for some variables, but as we use only \textit{soft} constraints, the solver may ignore some constraints if it cannot generate a satisfying assignment. 

The second part requires models from different \emph{equivalence classes}, where an equivalence class includes all the variables that are equal in the model. For example, if the model is $x_0 = x, x_1 = x$, where $x$ is a value of the corresponding type, then $x_0$ and $x_1$ belong to the same equivalence class. Considering equivalence classes is particularly important for variables of uninterpreted types; the solver cannot provide actual values for them, thus it assigns fresh, unconstrained variables. However, different fresh variables do not lead to diverse models. 

%% file: sections/extensions.tex
Next, we describe various extensions of our algorithm that enable complex proofs.
\paragraph{Combining multiple candidate terms.} In \algref{lst:Algorithm}, each candidate term is validated separately. To enable proofs that require multiple instantiations of the \emph{same} formula, we developed an extension that validates multiple triggering terms at the same time. In such cases, the algorithm returns a \emph{set of terms} that are necessary and sufficient to prove unsat. \figref{fig:Inconsistent SMT-COMP} presents a simple example from \SMTCOMP{} 2019 pending benchmarks \cite{SMT-COMP2019Pending}. The input $\inputFormula = F_0 \wedge F_1$ is unsatisfiable, as there does not exist an interpretation for the function $\mathtt{U}$ that satisfies all the constraints: $F_1$ requires $\mathtt{U}(\mathtt{s})$ to be $\mathtt{true}$; if $F_0$ is instantiated for $x_0 = \mathtt{s}$, the solver learns that $\mathtt{U}(\mathtt{il})$ must be $\mathtt{true}$ as well; however, if $x_0 = \mathtt{il}$, then $\mathtt{U}(\mathtt{il})$ must be $\mathtt{false}$, which is a contradiction. Exposing the inconsistency thus requires two instantiations of $F_0$, triggered by $\mathtt{f}(\mathtt{s})$ and $\mathtt{f}(\mathtt{il})$, respectively. We generate both triggering terms, but in separate iterations (independently, both fail to validate). However, by validating them simultaneously (i.e., conjoin both of them to $\inputFormula$), our algorithm identifies the required triggering term $T = \mathtt{dummy}(\mathtt{f}(\mathtt{s}), \mathtt{f}(\mathtt{il}))$.

\begin{figure}[t]
\begin{small}
\[
\begin{array}{ll} 
F_0\colon & \forall x_0\colon \mathrm{S} :: \{\mathtt{f}(x_0)\} \; \neg \mathtt{U}(x_0) \vee (\mathtt{U}(\mathtt{f}(x_0)) \wedge \mathtt{f}(x_0) = \mathtt{il} \wedge x_0 \neq \mathtt{il}) \\
F_1\colon & \mathtt{U}(\mathtt{s})  
\end{array}
\]
\end{small}
\vspace{-4mm}
\caption{Benchmark from SMT-COMP 2019 \cite{SMT-COMP2019Pending}. The formulas set contradictory constraints on the function $\mathtt{U}$. $\mathrm{S}$ is an uninterpreted type, $\mathtt{s}$ and $\mathtt{il}$ are user-defined constants of type $\mathrm{S}$. Synthesizing the triggering term $\mathtt{dummy}(\mathtt{f}(\mathtt{s}), \mathtt{f}(\mathtt{il}))$ requires multiple candidate terms. We use conjunctions here for simplicity, but our pre-processing applies distributivity of disjunction over conjunction and splits $F_0$ into three different formulas with unique names for the quantified variables. \label{fig:Inconsistent SMT-COMP}}
\end{figure}

\paragraph{Unification across multiple instantiations.} The clusters constructed by our algorithm are sets (see \algref{lst:Algorithm1}, line 12), so they contain a formula at most once, even if it is similar to multiple other formulas from the cluster. We thus consider the rewritings for multiple instantiations of the same formula separately, in different iterations. To handle cases that require multiple (but boundedly many) instantiations, we extend the algorithm with a parameter $\Phi$, which bounds the maximum frequency of a \textit{quantified} conjunct within the formulas $G$. That is, it allows a similar quantified formula, as well as $F$ itself, to be added to a cluster more than once (after performing variable renaming, to ensure that the names of the quantified variables are still globally unique). This results in an equisatisfiable formula for which our algorithm determines multiple triggering terms. Inputs whose unsatisfiability proofs require an unbounded number of instantiations  typically contain a matching loop, thus we do not consider them here. 

\begin{figure}
\begin{small}
\[
\begin{array}{ll} 
F_0\colon & \forall e_0\colon \mathrm{U} :: \{\mathtt{some}(e_0)\} \; \neg (\mathtt{some}(e_0) = \mathtt{none}) \\
F_1\colon & \forall op_1\colon \mathrm{U}, e_1\colon \mathrm{U} :: \{\mathtt{some}(e_1), \mathtt{get}(op_1)\} \; \neg (\mathtt{get}(op_1) = e_1) \vee (op_1 = \mathtt{some}(e_1)) \\
F_2\colon & \forall op_2\colon \mathrm{U}, e_2\colon \mathrm{U} :: \{\mathtt{some}(e_2),  \mathtt{get}(op_2)\} \; \neg (op_2 = \mathtt{some}(e_2)) \vee (\mathtt{get}(op_2) = e_2)
\end{array}
\]
\end{small}
\vspace{-4mm}
\caption{Fragment of Gobra's option types axiomatization. $\mathrm{U}$ is an uninterpreted type, $\mathtt{none}$ is a user-defined constant of type $\mathrm{U}$. $F_1 - F_2$ have \textit{multi-patterns} (\appendixref{sec:more_examples}). Synthesizing the triggering term $\mathtt{dummy}(\mathtt{some}(\mathtt{get}(\mathtt{none})))$ requires type-based constraints.\label{fig:Type-based constraints}}
\end{figure}

\paragraph{Type-based constraints.} The rewritings of the form $x_i = x_j$ can be too imprecise (especially for quantified variables of uninterpreted types), as they do not constrain the $rhs$. In \figref{fig:Type-based constraints}, the solver cannot provide concrete values of type $\mathrm{U}$ for $e_1$ and $op_1$, it can only assign fresh, unconstrained variables (e.g., $e$ and $op$). However, the triggering terms $\mathtt{some}(e)$ and $\mathtt{get}(op)$  are not sufficient to prove unsat; one would additionally need the rewriting $e_1 = \mathtt{get}(op_1)$, which cannot be identified by our unification from \secref{sec:groups}. To address such scenarios, we extend the unification to also consider as $rhs$ a constant or an uninterpreted function from the body of the similar formulas, which has the same type as the quantified variable from the left-hand side. For \figref{fig:Type-based constraints}, it will thus generate the rewritings $R = \{e_0 =  \mathtt{get}(op_2), e_1 =  \mathtt{get}(op_2), op_1 = \mathtt{none}, op_2 = \mathtt{none}\}$ (this is one of the alternatives). These type-based constraints allow us to synthesize the triggering term $T = \mathtt{dummy}(\mathtt{some}(\mathtt{get}(\mathtt{none})))$, which exposes the unsoundness from Gobra's option types axiomatization.

\paragraph{Unification for sub-terms.} \figref{fig:Inconsistent sub-terms} shows an example which cannot be solved by any extension discussed so far, since it requires semantic reasoning: by applying $\mathtt{f}$ on both sides of the equality, one can learn from $F_1$ that $\mathtt{f}(\mathtt{g}(2020)) = \mathtt{f}(\mathtt{g}(2021))$. From $F_0$ though, $\mathtt{f}(\mathtt{g}(2020)) = 2020$ and $\mathtt{f}(\mathtt{g}(2021)) = 2021$, which implies that $2020 = 2021$, i.e., $\mathtt{false}$. Our extended algorithm synthesizes the required triggering term $T = \mathtt{dummy}(\mathtt{f}(\mathtt{g}(2020)), \mathtt{f}(\mathtt{g}(2021)))$ by applying the unification also to \textit{sub-terms}. In \figref{fig:Inconsistent sub-terms}, trying to unify $\mathtt{f}(\mathtt{g}(x_0))$ does not produce any rewritings, as $F_1$ does not contain $\mathtt{f}(\mathtt{g})$. We thus unify the subterm $\mathtt{g}(x_0)$ with $\mathtt{g}(2020)$ and $\mathtt{g}(2021)$ and obtain the rewritings $R = \{ x_0 = 2020, x_0 = 2021\}$. Together with the extension for combining multiple candidate terms described above, these rewritings provide sufficient information for the unsatisfiability proof to succeed.

\begin{figure}[t]
\begin{small}
\[
\begin{array}{ll} 
F_0\colon & \forall x_0\colon \mathrm{Int} :: \{\mathtt{f}(\mathtt{g}(x_0))\} \; \mathtt{f}(\mathtt{g}(x_0)) = x_0 \\
F_1\colon & \mathtt{g}(2020) = \mathtt{g}(2021) \\
\end{array}
\]
\end{small}
\vspace{-4mm}
\caption{Formulas that set contradictory constraints on the function $\mathtt{f}$. Synthesizing the triggering term $\mathtt{dummy}(\mathtt{f}(\mathtt{g}(2020)), \mathtt{f}(\mathtt{g}(2021)))$ requires unification for sub-terms. \label{fig:Inconsistent sub-terms}}
\end{figure}

\paragraph{Alternative triggering terms.} Our algorithm returns the \textit{first} candidate term that successfully validates (\algref{lst:Algorithm}, line 18). However, it might also be useful to synthesize \textit{alternative} triggering terms for the same input, which may correspond to different completeness or soundness issues. Our tool provides this option and can also return \textit{all} the triggering terms found within the given timeout.

%% file: sections/additional_examples.tex
In this section, we illustrate our algorithm on various examples (including those from \figref{fig:motivation-boogie} and \figref{fig:Dafny seq}, and an example with nested quantifiers). We also explain how the algorithm supports quantifier-free formulas, synonym functions as patterns, multi-patterns and alternative patterns.

\paragraph{Nested quantifiers.} Our algorithm handles inputs with nested quantifiers as described in \secref{sec:groups}. We illustrate this aspect on the formulas from \figref{fig:Inconsistent list}, which axiomatize operations over lists of integers. The axioms $F_3$ and $F_4$ set contradictory constraints on $\mathtt{indexOf}$ when the element is not contained in the list. According to \algref{lst:Algorithm1}, one of the clusters generated for $F_3$ is $C=\{F_2, F_0\}$, with the rewritings $R=\{l_3 = l_2, el_3 = el_2, l_2 = l_0\}$. The algorithm then computes the instantiations for $F_0$ and $F_2$; as $F_2$ contains nested quantifiers, we remove both of them and obtain: $\instantiation{F_2} = \{\neg \mathtt{isEmpty}(l_2), \mathtt{isEmpty}(l_2) \wedge \neg \mathtt{has}(l_2, el_2)\}$, $\instantiation{F_0} = \{\neg (l_0 = \mathtt{EmptyList}), l_0 = \mathtt{EmptyList} \wedge \mathtt{isEmpty}(l_0) \}$.  The model of the corresponding $G$ formula and $R$ allow us to synthesize the required triggering term $T= \mathtt{dummy}(\mathtt{isEmpty}(\mathtt{EmptyList}), \mathtt{has}(\mathtt{EmptyList}, 0))$.

\begin{figure}[t]
\begin{small}
\[
\begin{array}{ll} 

F_0\colon & \forall l_0\colon \mathrm{L} :: \{\mathtt{isEmpty}(l_0)\} \; \neg (l_0 = \mathtt{EmptyList}) \vee \mathtt{isEmpty}(l_0) \\
F_1\colon & \forall l_1\colon \mathrm{L} :: \{\mathtt{isEmpty}(l_1)\} \; \mathtt{isEmpty}(l_1) \vee \mathtt{has}(l_1, \mathtt{f_1}(l_1)) \\
F_2\colon & \forall l_2\colon \mathrm{L} :: \{\mathtt{isEmpty}(l_2)\} \; \neg \mathtt{isEmpty}(l_2) \vee  
                               \forall el_2\colon \mathrm{Int} :: \{\mathtt{has}(l_2, el_2)\} \; \neg \mathtt{has}(l_2, el_2) \\
F_3\colon & \forall l_3\colon \mathrm{L}, el_3\colon \mathrm{Int} :: \{\mathtt{has}(l_3, el_3)\} \; \mathtt{has}(l_3, el_3) \vee (\mathtt{indexOf}(l_3, el_3) = -1) \\
F_4\colon & \forall l_4\colon \mathrm{L}, el_4\colon \mathrm{Int} :: \{\mathtt{indexOf}(l_4, el_4)\} \; \mathtt{indexOf}(l_4, el_4) \geq 0 
\end{array}
\]
\end{small}
\vspace{-4mm}
\caption{Formulas that set contradictory constraints on $\mathtt{indexOf}$. $\mathrm{L}$ is an uninterpreted type, $\mathtt{EmptyList}$ is a user-defined constant of type $\mathrm{L}$. $\mathtt{f_1}$ is a Skolem function, which replaces a nested existential quantifier. $F_2$ contains nested universal quantifiers. \label{fig:Inconsistent list}}
\end{figure}

\paragraph{Quantifier-free formulas.} 
Our algorithm from~\algref{lst:Algorithm} iterates only over quantified conjuncts, but leverages the additional information provided by quantifier-free formulas and includes them in the clusters even if the unification cannot find a rewriting (\algref{lst:Algorithm1}, line 9). Since quantifier-free conjuncts can be seen as already instantiated formulas, we only have to cover all their disjuncts (\algref{lst:Algorithm}, line 7).

\begin{figure}[b]
\begin{small}
\[
\begin{array}{ll} 
F_0\colon & \forall x_0\colon \mathrm{Int} :: \{\mathtt{len}(\mathtt{nxt}(x_0)\} \; \mathtt{len}(x_0) > 0 \\
F_1\colon & \forall x_1\colon \mathrm{Int} :: \{\mathtt{len}(\mathtt{nxt}(x_1)\} \; (\mathtt{nxt}(x_1) = x_1) \vee (\mathtt{len}(x_1) = \mathtt{len}(\mathtt{nxt}(x_1)) + 1) \\
F_2\colon & \forall x_2\colon \mathrm{Int} :: \{\mathtt{len}(\mathtt{nxt}(x_2)\} \; \neg (\mathtt{nxt}(x_2) = x_2) \vee (\mathtt{len}(x_2) = 1) \\
F_3\colon & \mathtt{len}(7) \leq 0
\end{array}
\]
\end{small}
\vspace{-4mm}
\caption{Boogie example from~\figref{fig:motivation-boogie} encoded in our input format. $F_0$--$F_2$ represent the axiom, while the quantifier-free formula $F_3$ is the negation of the assertion (verifiers discharge their proof obligations by showing that the negation is unsatisfiable).\label{fig:Inconsistent Boogie}}
\end{figure}

\paragraph{Boogie example.} \figref{fig:Inconsistent Boogie} shows the example from \figref{fig:motivation-boogie} encoded in our format. The quantifier-free formula $F_3$ (i.e., the verification condition) is similar to $F_0$ (they share the symbol $\mathtt{len}$) and unifies through the rewritings $R = \{x_0 = 7\}$. We obtain the required triggering term $T = \mathtt{dummy}(\mathtt{len}(\mathtt{nxt}(7)))$ from the model of the formula $G = \neg F_0' \wedge \instantiations{F_3}{0} \wedge \bigwedge{R} = (\mathtt{len}(x_0) \leq 0) \wedge (\mathtt{len}(7) \leq 0) \wedge (x_0 = 7)$.

\paragraph{Dafny example.} Our algorithm can synthesize various triggering terms that expose the unsoundness from \figref{fig:Dafny seq}, depending on the values of its parameters. We explain here one, obtained for $\sigma = 0.1$. For $\code{depth} = 0$, the algorithm checks each formula $F_0$ -- $F_4$ in isolation. As they are all individually satisfiable, it continues with $\code{depth} = 1$. Due to space constraints, we only present the iteration for $F_3$.

$F_3$ shares at least two uninterpreted symbols with each of the other formulas, so there are various alternative rewritings: $s_3 = \mathtt{Empty}(t_1)$, $s_3 = s_4$, $s_3 = \mathtt{Build}(s_4, i_4, v_4, l_4)$, etc. As we consider clusters-rewritings pairs in which each quantified variable has maximum one rewriting, one such pair is $(C = \{F_4\}, R = \{s_3 = \mathtt{Build}(s_4, i_4, v_4, l_4)\})$. $F_4$ has two disjuncts, so its instantiations are: $\instantiation{F_4} = \{\mathtt{typ}(s_4) = \mathtt{Type}(\mathtt{typ}(v_4),$ $\neg(\mathtt{type}(s_4) = \mathtt{Type}(\mathtt{typ}(v_4)) \wedge \mathtt{Len}(\mathtt{Build}(s_4, i_4, v_4, l_4)) = l_4\}$. From these instantiations and the rewritings $R$, we derive two formulas: $G_0  = \neg F_3' \wedge \instantiations{F_4}{0} \wedge \bigwedge R$, with the model $s_3 = s$, $s_4 = s'$, $i_4 = 0$, $v_4 = v$, $l_4 = 1$ and $G_1  = \neg F_3' \wedge \instantiations{F_4}{1} \wedge \bigwedge R$, with the model $s_3 = s, s_4 = s', i_4 = 0, v_4 = v, l_4 = -1$, where $s$, $s'$, and $v$ are fresh variables of type $\mathrm{U}$ (We use indexes for the $G$ formulas to refer to them later). We then construct the candidate triggering terms from the patterns of the formulas $F_3$ and $F_4$. We replace $s_3$ by its right hand side in the rewriting, \ie{}~$\mathtt{Build}(s_4, i_4, v_4, l_4)$, and all the other quantified variables by their constants from the model. The result after removing redundant terms is: $T_0 = \mathtt{dummy}(\mathtt{Len}(\mathtt{Build}(t, 0, v, 1)))$ and $T_1 = \mathtt{dummy}(\mathtt{Len}(\mathtt{Build}(t, 0, v, -1)))$. Since the validation step fails for both $T_0$ and $T_1$, we continue with the other $(C, R)$ pairs, the remaining quantified conjuncts and their similarity clusters. 

If no candidate term is sufficient to prove unsat, our algorithm expends the clusters. To scale to real-world axiomatizations, it efficiently reuses the results from the previous iterations; i.e., it prunes the search space if a previously synthesized formula $G$ is unsatisfiable and it strengthens $G$ if it is satisfiable. The pair $(C = \{F_4\}, R = \{s_3 = \mathtt{Build}(s_4, i_4, v_4, l_4)\})$ can be extended to $(C = \{F_4, F_1\}, R = \{s_3 = \mathtt{Build}(s_4, i_4, v_4, l_4), s_4 = \mathtt{Empty}(t_1), t_1 = \mathtt{typ}(v_4)\})$, as $F_1$ is similar to $F_4$ through the rewritings $R = \{s_4 = \mathtt{Empty}(t_1), t_1 = \mathtt{typ}(v_4)\}$. We thus conjoin the instantiation of $F_1$ and the two additional rewritings to the formulas $G_0$ and $G_1$ from the previous iteration. This is equivalent to recomputing the similarity cluster, the rewritings, and the combinations of instantiations. We then obtain: $G_0' = G_0 \wedge (\mathtt{type}(\mathtt{Empty}(t_1)) = \mathtt{Type}(t_1)) \wedge (s_4 = \mathtt{Empty}(t_1)) \wedge (t_1 = \mathtt{typ}(v_4))$, which is unsatisfiable, and $G_1' = G_1 \wedge (\mathtt{type}(\mathtt{Empty}(t_1)) = \mathtt{Type}(t_1)) \wedge (s_4 = \mathtt{Empty}(t_1)) \wedge (t_1 = \mathtt{typ}(v_4))$ with the model: $s_3 = s$, $s_4 = s'$, $i_4 = 0$, $v_4 = v$, $l_4 = -1$, $t_1 = t$, where $s$, $s'$, $v$, and $t$ are fresh variables of types $\mathrm{U}$ and $\mathrm{V}$. From this model and the rewritings we construct the triggering term $T = \mathtt{dummy}(\mathtt{Len}(\mathtt{Build}(\mathtt{Empty}(\mathtt{typ}(v)), 0, v, -1)))$, which is sufficient to expose the inconsistency between $F_3$ and $F_4$. 

\begin{figure}
\begin{small}
\[
\begin{array}{ll}
F\colon & \forall a\colon \mathrm{Int}, b\colon \mathrm{Int}, size\colon \mathrm{Int} :: \{\mathtt{both\_ptr}(a, b, size)\} \\
& \hspace{0.6cm} \mathtt{both\_ptr}(a, b, size) * size \leq a - b  \\
\end{array}
\]
\end{small}
\vspace{-4mm}
\caption{Inconsistent formula from a VCC/HAVOC~\cite{VCC,Shaunak2007} benchmark from SMT-COMP \cite{SMT-COMP2020}, which cannot be proved unsat by MBQI. Our synthesized triggering term $\mathtt{dummy}(\mathtt{both\_ptr}(-2, -1, 0))$ allows E-matching to refute the formula. \label{fig:one_axiom}}
\end{figure}

\paragraph{VCC/HAVOC example.} \figref{fig:one_axiom} presents a fragment of a benchmark which could be solved by our algorithm, but could not be proved by MBQI. $F$, which was part of a set of 160 formulas, is inconsistent by itself: when $size = 0$, E-matching can refute it for any integer values $a$, $b$, such that $a \leq b$. Our algorithm synthesizes the required triggering term in $\approx$7s because it initially considers each quantified conjunct in isolation. The formula $G = \neg F' = \mathtt{both\_ptr}(a, b, size) * size >  a - b$ is satisfiable and the simplest models the solver can provide (without assigning an interpretation to the uninterpreted function $\mathtt{both\_ptr}$) all include $size = 0$.

\begin{figure}[b]
\begin{small}
\[
\begin{array}{ll} 
F_0\colon & \forall s_0\colon \mathrm{ISeq}, l_0\colon \mathrm{Int}, h_0\colon \mathrm{Int} :: \{\mathtt{sum}(s_0, l_0, h_0)\} \; \mathtt{sum}(s_0, l_0, h_0) =  \mathtt{sum\_syn}(s_0, l_0, h_0)\\

F_1\colon & \forall s_1\colon \mathrm{ISeq}, l_1\colon \mathrm{Int}, h_1\colon \mathrm{Int} :: \{\mathtt{sum}(s_1, l_1, h_1)\} \; \neg (l_1 \geq h_1) \vee (\mathtt{sum\_syn}(s_1, l_1, h_1) = 0)\\

F_2\colon & \forall s_2\colon \mathrm{ISeq}, l_2\colon \mathrm{Int}, h_2\colon \mathrm{Int} :: \{\mathtt{sum}(s_2, l_2, h_2)\} \; \neg (l_2 \leq h_2) \; \vee \\

& \hspace{0.8cm} (\mathtt{sum\_syn}(s_2, l_2, h_2) =  \mathtt{sum\_syn}(s_2, l_2 + 1, h_2) + \mathtt{seq.nth}(s_2, l_2))\\

F_3\colon & \mathtt{seq.nth}(\mathtt{empty}, 0) = -1
\end{array}
\]
\end{small}
\vspace{-4mm}
\caption{Formulas with synonym functions as patterns that axiomatize sequence comprehensions and set contradictory constraints on the function $\mathtt{sum\_syn}$. $\mathrm{ISeq}$ is a user-defined type, $\mathtt{empty}$ is a user-defined constant of type $\mathrm{ISeq}$ (i.e., the empty sequence). \label{fig:Inconsistent synonym functions}}
\end{figure}

\paragraph{Synonym functions as patterns.} For the examples discussed so far, the functions used as patterns were also present in the body of the quantifiers. However, to have a better control over the instantiations, one can also write formulas where the patterns are additional uninterpreted functions, which do not appear in the bodies. Such patterns are not uncommon in proof obligations. \figref{fig:Inconsistent synonym functions} shows an example, which uses the synonym functions technique \cite{Leino2009} to avoid matching loops. $\mathtt{sum}$ and $\mathtt{sum\_syn}$ compute the sum of the elements of a sequence, between a lower and an upper bound. The two functions are identical (according to $F_0$), but only $\mathtt{sum}$ is used as a pattern. For equal bounds, $F_1$ and $F_2$ set contradictory constraints on the interpretation of $\mathtt{sum\_syn}$. $\mathtt{seq.nth}$ returns the n-th element of the sequence. Using the information from the quantifier-free formula $F_3$, our algorithm generates the triggering term $T=\mathtt{dummy}(\mathtt{sum}(\mathtt{empty}, 0, 0), \mathtt{sum}(\mathtt{empty}, 0+1, 0))$. The term "$0+1$" comes from the rewriting $l_0 = l_2 +1$. Addition is a built-in function, but is used as an argument to the uninterpreted function $\mathtt{sum\_syn}$, thus, it is supported by our unification. Our algorithm is syntactic, so it does not perform arithmetic operations, it just substitutes $l_2$ with its value from the model. The solver then performs theory reasoning and concludes unsat.

\paragraph{Multi-patterns and alternative patterns.} SMT solvers allow patterns to contain multiple terms, all of which must be present to perform an instantiation. $F_1$ in \figref{fig:Inconsistent multi-patterns} has such a \emph{multi-pattern} and can be instantiated only when triggering terms that match both $\{\mathtt{g}(b_1)\}$ \emph{and} $\{\mathtt{f}(x_1)\}$ are present in the SMT run. Our algorithm directly supports multi-patterns, as the procedure $\algCandidateTerm$ instantiates all the patterns from the given cluster (see \algref{lst:Algorithm2}, line 9). For the example from \figref{fig:Inconsistent multi-patterns}, our technique synthesizes the triggering term $T = \mathtt{dummy}(\mathtt{f}(7), \mathtt{g}(b))$ from the rewritings $R=\{x_0 = x_1\}$ and the model of the formula $G = \neg F_0' \wedge \instantiations{F_1}{1} \wedge \bigwedge R = (\mathtt{f}(x_0) = 7) \wedge  (\neg \mathtt{g}(b_1) \wedge \mathtt{f}(x_1) = x_1) \wedge (x_0 = x_1)$. $b$ is a fresh, unconstrained variable of the uninterpreted type $\mathrm{B}$.

\begin{figure}[t]
\begin{small}
\[
\begin{array}{ll} 
F_0\colon & \forall x_0\colon \mathrm{Int} :: \{\mathtt{f}(x_0)\} \; \mathtt{f}(x_0) \neq 7 \\
F_1\colon & \forall b_1\colon \mathrm{B}, x_1\colon \mathrm{Int} :: \{\mathtt{g}(b_1), \mathtt{f}(x_1)\} \; \mathtt{g}(b_1) \vee (\mathtt{f}(x_1) = x_1) \\
F_2\colon & \forall b_2\colon \mathrm{B} :: \{\mathtt{g}(b_2)\} \; \neg \mathtt{g}(b_2)  
\end{array}
\]
\end{small}
\vspace{-4mm}
\caption{Formulas that set contradictory constraints on the function $\mathtt{f}$. $F_1$ has a multi-pattern. $\mathrm{B}$ is an uninterpreted type. \label{fig:Inconsistent multi-patterns}}
\end{figure}

Formulas can also contain \emph{alternative patterns}. For example, the quantified formula $\forall x\colon \mathrm{Int} :: \{\mathtt{f}(x)\} \: \{\mathtt{h}(x)\} \; \mathtt{f}(x) \neq 7 \vee (\mathtt{h}(x) = 6)$ is instantiated only if there exists a triggering term that matches $\{\mathtt{f}(x)\}$ \emph{or} one that matches $\{\mathtt{h}(x)\}$. Our algorithm does not differentiate between multi-patterns and alternative patterns, thus it always synthesizes the arguments for \textit{all} the patterns of a cluster. For alternative patterns, this results in an over-approximation of the set of necessary triggering terms. However, the minimization step (performed before returning the triggering term that successfully validates), removes the unnecessary terms. 

%% file: sections/optimizations.tex
In this section, we present various optimizations implemented in our tool, which allow the algorithm to scale to real-world verification benchmarks.

\paragraph{Grammar.}
The grammar from \figref{fig:Grammar} allows us to simplify the presentation of the algorithm. However, eliminating conjunctions by applying distributivity and splitting (as described in \secref{sec:input}) can result in an exponential increase in the number of terms and introduce redundancy, affecting the performance. Conjunction elimination is not implemented in Z3's \code{NNF} tact, thus it is not performed automatically. We apply this transformation only at the top-level, i.e., we do not recursively distribute disjunctions over conjunctions. For this reason, the input conjuncts $F$ supported by our tool can actually contain conjunctions, in which case we use an extended algorithm when computing the instantiations, to ensue that all the resulting $G$ formulas are still quantifier-free. The number of conjuncts and the number of quantifiers reported in \tabref{tab:verifiers} and \tabref{tab:smt_comp} were computed \textit{before} applying distributivity, thus they are not artificially increased. 

\paragraph{Rewritings.} The restrictive shapes of our rewritings (from \secref{sec:groups}, step~2), ensure that their number is finite, because if it exists, the most general unifier is unique up to variable renaming, i.e., substitutions of the type $\{x_i \rightarrow x_j, x_j \rightarrow x_i\}$~\cite{Baader2001}. (Such substitutions are rewritings of the shapes (1) and (2) where $rhs$ is also a quantified variable.) However, for most practical examples, the number of rewritings is very large, thus our implementation identifies them lazily, in increasing order of cardinality. If a rewriting $r \in R$ leads to an unsat $G$ formula for some instantations, then we discard all the subsequent $G$ formulas that contain $r$ and the same instantiations (they will also be unsatisfiable). To make sure that the algorithm terminates within a given amount of time, in our experiments we bound the number of $G$ formulas derived for each quantified conjunct $F$ to 100. 

\paragraph{Instantiations.} Our implementation computes lazily the Cartesian product of the instantiations (\algref{lst:Algorithm}, line 9), since it can also have a high number of elements. However, many of them are in practice unsatisfiable, thus our tool efficiently identifies trivial conflicts (e.g., $\neg D_i \wedge D_i$), pruning the search space accordingly.

\paragraph{Candidate terms.} To improve the performance of our algorithm, we keep track of all the candidate triggering terms that failed to validate (i.e., of the models from which they were synthesized). Then, we add constraints (similar to the conjunct $\neg\mathtt{model}$ from \algref{lst:Algorithm}, line 19) to ensure the solver does not provide previously-seen models for the quantified variables from the same set of patterns.

%% file: sections/benchmarks_smt_comp.tex
Next, we describe our filtering step, for identifying files with triggering issues.

\medskip
We collected 27,716 benchmarks from \SMTCOMP{} 2020~\cite{SMT-COMP2020} (single query track), with ground truth \emph{unsat} and at least one pattern (as this suggests they were designed for E-matching). We then ran Z3 to infer the missing patterns and to transform the formulas into NNF and removed all benchmarks for which the inference or the transformation did not succeed within 600s per file and 4s per formula. We also removed the files with features not yet supported by PySMT~\cite{pysmt2015}, the parsing library used in our experiments (e.g., sort signatures in datatypes declarations, but we did extend PySMT to handle, e.g., patterns and overloaded functions). This filtering resulted in 6,481 benchmarks. We then ran E-matching and kept only those examples that could not be solved within 600s due to incompleteness in instantiating quantifiers (our work only targets this incompleteness, but the SMT-COMP suite also contains other solving challenges). We thus obtained 61 files from \SPECSHARP~\cite{SpecSharp}, VCC~\cite{VCC}, Havoc~\cite{Shaunak2007}, Simplify~\cite{Detlefs2005}, and the Bit-Width-Independent (BWI) encoding~\cite{Niemetz2019}, summarized in \tabref{tab:smt_comp}.

%% file: paper.bbl
\begin{thebibliography}{10}
\providecommand{\url}[1]{\texttt{#1}}
\providecommand{\urlprefix}{URL }
\providecommand{\doi}[1]{https://doi.org/#1}

\bibitem{ArrayMax}
Array maximum, by elimination (2021),
  \url{http://viper.ethz.ch/examples/max-array-elimination.html}

\bibitem{Fstar1848}
F* issue 1848 (2021), \url{https://github.com/FStarLang/FStar/issues/1848}

\bibitem{ViperTestSuite}
Viper test suite (2021),
  \url{https://github.com/viperproject/silver/tree/master/src/test/resources}

\bibitem{Amighi2016}
Amighi, A., Blom, S., Huisman, M.: Vercors: {A} layered approach to practical
  verification of concurrent software. In: {PDP}. pp. 495--503. {IEEE} Computer
  Society (2016), \url{https://ieeexplore.ieee.org/abstract/document/7445381}

\bibitem{Astrauskas19}
Astrauskas, V., M\"uller, P., Poli, F., Summers, A.J.: Leveraging {R}ust types
  for modular specification and verification. In: Object-Oriented Programming
  Systems, Languages, and Applications (OOPSLA). vol.~3, pp. 147:1--147:30. ACM
  (2019). \doi{10.1145/3360573}

\bibitem{Baader2001}
Baader, F., Snyder, W.: Unification theory. In: Robinson, J.A., Voronkov, A.
  (eds.) Handbook of Automated Reasoning. pp. 445--532. Elsevier and MIT Press
  (2001)

\bibitem{boogie}
Barnett, M., Chang, B.Y.E., DeLine, R., Jacobs, B., Leino, K.R.M.: Boogie: A
  modular reusable verifier for object-oriented programs. In: de~Boer, F.S.,
  Bonsangue, M.M., Graf, S., de~Roever, W.P. (eds.) Formal Methods for
  Components and Objects (FMCO). Lecture Notes in Computer Science, vol.~5, pp.
  364--387. Springer (2005)

\bibitem{SpecSharp}
Barnett, M., F\"ahndrich, M., Leino, K.R.M., M\"uller, P., Schulte, W., Venter,
  H.: Specification and verification: The {Spec\#} experience. Communications
  of the ACM  \textbf{54}(6),  81--91 (June 2011)

\bibitem{Barrett2011}
Barrett, C., Conway, C.L., Deters, M., Hadarean, L., Jovanovi{\'{c}}, D., King,
  T., Reynolds, A., Tinelli, C.: Cvc4. In: Gopalakrishnan, G., Qadeer, S.
  (eds.) Computer Aided Verification. pp. 171--177. Springer Berlin Heidelberg,
  Berlin, Heidelberg (2011)

\bibitem{Barrett2017}
Barrett, C., Fontaine, P., Tinelli, C.: {The SMT-LIB Standard: Version 2.6}.
  Tech. rep., Department of Computer Science, The University of Iowa (2017),
  available at {\tt www.SMT-LIB.org}

\bibitem{Shaunak2007}
Chatterjee, S., Lahiri, S.K., Qadeer, S., Rakamari{\'{c}}, Z.: A reachability
  predicate for analyzing low-level software. In: Grumberg, O., Huth, M. (eds.)
  Tools and Algorithms for the Construction and Analysis of Systems. pp.
  19--33. Springer Berlin Heidelberg, Berlin, Heidelberg (2007)

\bibitem{DarvasLeino07}
Darvas, A., Leino, K.R.M.: Practical reasoning about invocations and
  implementations of pure methods. In: Dwyer, M.B., Lopes, A. (eds.)
  Fundamental Approaches to Software Engineering (FASE). LNCS, vol.~4422, pp.
  336--351. Springer-Verlag (2007)

\bibitem{Detlefs2005}
Detlefs, D., Nelson, G., Saxe, J.B.: Simplify: A theorem prover for program
  checking. J. ACM  \textbf{52}(3),  365--473 (May 2005).
  \doi{10.1145/1066100.1066102},
  \url{http://doi.acm.org/10.1145/1066100.1066102}

\bibitem{Eilers18}
Eilers, M., M\"uller, P.: Nagini: A static verifier for python. In: Chockler,
  H., Weissenbacher, G. (eds.) Computer Aided Verification (CAV). LNCS, vol.
  10982, pp. 596--603. Springer International Publishing (2018).
  \doi{10.1007/978-3-319-96145-3\_33}

\bibitem{pysmt2015}
Gario, M., Micheli, A.: {PySMT}: a solver-agnostic library for fast prototyping
  of {SMT-based} algorithms. In: SMT Workshop 2015 (2015)

\bibitem{MBQI}
Ge, Y., de~Moura, L.: Complete instantiation for quantified formulas in
  satisfiabiliby modulo theories. In: Bouajjani, A., Maler, O. (eds.) Computer
  Aided Verification. pp. 306--320. Springer Berlin Heidelberg, Berlin,
  Heidelberg (2009)

\bibitem{HeuleKassiosMuellerSummers13}
Heule, S., Kassios, I.T., M{\"u}ller, P., Summers, A.J.: Verification condition
  generation for permission logics with abstract predicates and abstraction
  functions. In: Castagna, G. (ed.) European Conference on Object-Oriented
  Programming (ECOOP). Lecture Notes in Computer Science, vol.~7920, pp.
  451--476. Springer (2013)

\bibitem{Kovacs2013}
Kov{\'a}cs, L., Voronkov, A.: First-order theorem proving and {Vampire}. In:
  Sharygina, N., Veith, H. (eds.) Computer Aided Verification. pp. 1--35.
  Springer Berlin Heidelberg, Berlin, Heidelberg (2013)

\bibitem{Leino2010}
Leino, K.R.M.: Dafny: An automatic program verifier for functional correctness.
  In: Clarke, E.M., Voronkov, A. (eds.) Logic for Programming, Artificial
  Intelligence, and Reasoning. pp. 348--370. Springer Berlin Heidelberg,
  Berlin, Heidelberg (2010)

\bibitem{Leino2009}
Leino, K.R.M., Monahan, R.: Reasoning about comprehensions with first-order
  {SMT} solvers. In: Proceedings of the 2009 ACM Symposium on Applied
  Computing. pp. 615--622. SAC'09, Association for Computing Machinery, New
  York, NY, USA (2009). \doi{10.1145/1529282.1529411},
  \url{https://doi.org/10.1145/1529282.1529411}

\bibitem{LeinoMueller08}
Leino, K.R.M., M\"uller, P.: Verification of equivalent-results methods. In:
  Drossopoulou, S. (ed.) European Symposium on Programming (ESOP). Lecture
  Notes in Computer Science, vol.~4960, pp. 307--321. Springer-Verlag (2008)

\bibitem{LeinoRummer2010}
Leino, K.R.M., R{\"u}mmer, P.: A polymorphic intermediate verification
  language: Design and logical encoding. In: Esparza, J., Majumdar, R. (eds.)
  Tools and Algorithms for the Construction and Analysis of Systems. pp.
  312--327. Springer Berlin Heidelberg, Berlin, Heidelberg (2010)

\bibitem{Moskal09}
Moskal, M.: Programming with triggers. In: {{S}{M}{T}}. ACM International
  Conference Proceeding Series, vol.~375, pp. 20--29. {ACM} (2009)

\bibitem{Z3}
de~Moura, L., Bj{\o}rner, N.: Z3: An efficient {SMT} solver. In: Ramakrishnan,
  C.R., Rehof, J. (eds.) Tools and Algorithms for the Construction and Analysis
  of Systems. pp. 337--340. Springer Berlin Heidelberg, Berlin, Heidelberg
  (2008)

\bibitem{MuellerSchwerhoffSummers16}
M{\"u}ller, P., Schwerhoff, M., Summers, A.J.: Viper: A verification
  infrastructure for permission-based reasoning. In: Jobstmann, B., Leino,
  K.R.M. (eds.) Verification, Model Checking, and Abstract Interpretation
  (VMCAI). LNCS, vol.~9583, pp. 41--62. Springer-Verlag (2016)

\bibitem{Niemetz2019}
Niemetz, A., Preiner, M., Reynolds, A., Zohar, Y., Barrett, C., Tinelli, C.:
  Towards bit-width-independent proofs in {SMT} solvers. In: Fontaine, P. (ed.)
  Automated Deduction -- CADE 27. pp. 366--384. Springer International
  Publishing, Cham (2019)

\bibitem{Reger2016}
Reger, G., Bjorner, N., Suda, M., Voronkov, A.: {AVATAR} modulo theories. In:
  Benzm\"uller, C., Sutcliffe, G., Rojas, R. (eds.) GCAI 2016. 2nd Global
  Conference on Artificial Intelligence. EPiC Series in Computing, vol.~41, pp.
  39--52. EasyChair (2016). \doi{10.29007/k6tp},
  \url{https://easychair.org/publications/paper/7}

\bibitem{Reynolds2018}
Reynolds, A., Barbosa, H., Fontaine, P.: Revisiting enumerative instantiation.
  In: Beyer, D., Huisman, M. (eds.) Tools and Algorithms for the Construction
  and Analysis of Systems. pp. 112--131. Springer International Publishing,
  Cham (2018)

\bibitem{Reynolds2015}
Reynolds, A., Deters, M., Kuncak, V., Tinelli, C., Barrett, C.:
  Counterexample-guided quantifier instantiation for synthesis in {SMT}. In:
  Kroening, D., P{\u{a}}s{\u{a}}reanu, C.S. (eds.) Computer Aided Verification.
  pp. 198--216. Springer International Publishing, Cham (2015)

\bibitem{RudichDarvasMueller08}
Rudich, A., \'{A}d\'{a}m Darvas, M\"uller, P.: Checking well-formedness of
  pure-method specifications. In: Cuellar, J., Maibaum, T. (eds.) Formal
  Methods (FM). Lecture Notes in Computer Science, vol.~5014, pp. 68--83.
  Springer-Verlag (2008)

\bibitem{Rummer2012}
R{\"u}mmer, P.: E-matching with free variables. In: Bj{\o}rner, N., Voronkov,
  A. (eds.) Logic for Programming, Artificial Intelligence, and Reasoning. pp.
  359--374. Springer Berlin Heidelberg, Berlin, Heidelberg (2012)

\bibitem{VCC}
Schulte, W.: {VCC}: Contract-based modular verification of concurrent c. In:
  31st International Conference on Software Engineering, ICSE 2009. IEEE
  Computer Society (January 2008),
  \url{https://www.microsoft.com/en-us/research/publication/vcc-contract-based-modular-verification-of-concurrent-c/}

\bibitem{SMT-COMP2019Pending}
{SMT-COMP 2019}: The 14th international satisfiability modulo theories
  competition (including pending benchmarks) (2019),
  \url{https://smt-comp.github.io/2019/,
  https://clc-gitlab.cs.uiowa.edu:2443/SMT-LIB-benchmarks-tmp/benchmarks-pending}

\bibitem{SMT-COMP2020}
{SMT-COMP 2020}: The 15th international satisfiability modulo theories
  competition (2020), \url{https://smt-comp.github.io/2020/}

\bibitem{Sut16}
Sutcliffe, G.: {The CADE ATP System Competition - CASC}. AI Magazine
  \textbf{37}(2),  99--101 (2016)

\bibitem{Swamy2016}
Swamy, N., Hri\textcommabelow{t}cu, C., Keller, C., Rastogi, A.,
  Delignat-Lavaud, A., Forest, S., Bhargavan, K., Fournet, C., Strub, P.Y.,
  Kohlweiss, M., Zinzindohoue, J.K., Zanella-B\'{e}guelin, S.: Dependent types
  and multi-monadic effects in {F*}. In: Proceedings of the 43rd Annual ACM
  SIGPLAN-SIGACT Symposium on Principles of Programming Languages. pp.
  256--270. POPL '16, Association for Computing Machinery, New York, NY, USA
  (2016). \doi{10.1145/2837614.2837655},
  \url{https://doi.org/10.1145/2837614.2837655}

\bibitem{Swamy2013}
Swamy, N., Weinberger, J., Schlesinger, C., Chen, J., Livshits, B.: Verifying
  higher-order programs with the {Dijkstra} monad. In: Proceedings of the 34th
  annual ACM SIGPLAN conference on Programming Language Design and
  Implementation. pp. 387--398. PLDI '13 (2013),
  \url{https://www.microsoft.com/en-us/research/publication/verifying-higher-order-programs-with-the-dijkstra-monad/}

\bibitem{Voronkov2014}
Voronkov, A.: {AVATAR}: The architecture for first-order theorem provers. In:
  Biere, A., Bloem, R. (eds.) Computer Aided Verification. pp. 696--710.
  Springer International Publishing, Cham (2014)

\bibitem{Gobra}
Wolf, F.A., Arquint, L., Clochard, M., Oortwijn, W., Pereira, J.C., M\"uller,
  P.: {G}obra: Modular specification and verification of go programs. In:
  Silva, A., Leino, K.R.M. (eds.) Computer Aided Verification (CAV). pp.
  367--379. Springer International Publishing (2021)

\end{thebibliography}
